\documentclass[12pt]{article}
\usepackage{amssymb,amsmath}
\usepackage{latexsym}
\usepackage{graphicx}
\usepackage{psfrag}

\usepackage{fullpage}

\numberwithin{equation}{section}

\renewcommand{\theequation}{\arabic{section}.\arabic{equation}}

\begin{document}

\begin{titlepage}

\begin{flushright}
DFPD-10/TH22
\end{flushright}

\bigskip
\bigskip
\centerline{\Large \bf U-duality and non-BPS solutions}
\bigskip
\vspace{1cm}
\bigskip
\centerline{{\bf Gianguido Dall'Agata$^{1,2}$, Stefano Giusto$^{3,4}$ and Cl\'{e}ment Ruef$^{\, 5}$}}
\bigskip
\centerline{$^1$ Dipartimento di Fisica ``Galileo Galilei''}
\centerline{Universit\`a di Padova, Via Marzolo 8, 35131 Padova, Italy}
\bigskip
\centerline{$^2$INFN, Sezione di Padova,}
\centerline{Via Marzolo 8, 35131 Padova, Italy}
\bigskip
\centerline{$^3$ Dipartimento di Fisica}
\centerline{Universit\`a di Genova, via Dodecaneso, 33, 16146 Genova, Italy}
\bigskip
\centerline{$^4$INFN, Sezione di Genova,}
\centerline{via Dodecaneso, 33, 16146 Genova, Italy}
\bigskip
\centerline{$^5$ Max Planck Institute for Gravitation, Albert Einstein Institute}
\centerline{Am M\"uhlenberg 1, 14476 Golm, Germany}
\bigskip
\bigskip
\bigskip

\begin{abstract}

We derive the explicit action of the U-duality group of the $STU$ model on both BPS and non-BPS extremal multi-center solutions. As the class of known non-BPS extremal solutions is not closed under U-duality, we generate in this way new solutions. These should represent the most general class of extremal non-BPS multi-center under-rotating solutions of the $STU$ model.
\end{abstract}

\end{titlepage}

\tableofcontents

\section{Introduction}

The physics of  supersymmetric multicenter black hole solutions in four dimensions is surprisingly rich. Besides allowing one to study the decay of states along lines of marginal stability in the moduli space \cite{Denef:2000nb}, or to understand the quantization of spacetimes \cite{deBoer:2008zn}, some of these solutions descend from five-dimensional smooth horizonless solutions \cite{Bena:2004de,Berglund:2005vb} that have the same charges and mass as a black hole, and thus provide prime candidate for ``microstate geometries'' of extremal BPS black holes. Furthermore, as it has been becoming clear over the past few years, in the vicinity of the five-dimensional smooth solutions obtained by the uplift of four-dimensional multicenter solutions there exists an infinite-dimensional family of smooth horizonless solutions parameterized by several arbitrary functions of one variable \cite{Bena:2010gg}, whose quantization may yield an entropy that has the same parametric dependence on charges as that of black holes, and thus establish that Mathur's conjecture \cite{Mathur:2005zp, Bena:2007kg, Skenderis:2008qn, Chowdhury:2010ct} applies to extremal BPS black holes.

It is important to understand how much of the beautiful physics of BPS multicenter solutions extends to non-BPS multicenter solutions, and it is clear that none of the issues discussed above can be satisfactorily addressed in the absence of explicit supergravity solutions. Unfortunately, constructing such solutions is no easy task, essentially because even the simplest multicenter solutions depend on functions of two variables, and solving the underlying second-order Einstein's equations from scratch is essentially impossible in the absence of supersymmetry, or of another guiding principle. 

This formidable problem was therefore left untouched, until it was noted that also non-BPS single center extremal configurations allow for a first-order description both in four \cite{Ceresole:2007wx} and in five dimensions \cite{Lopes Cardoso:2007ky} and that, by changing a few signs in the equations that govern five-dimensional supersymmetric solutions \cite{Gauntlett:2002nw,Gutowski:2004yv,Bena:2004de}, one can obtain a set of first-order equations that govern a certain class of nonsupersymmetric extremal multicenter solutions \cite{Goldstein:2008fq}. 
The supersymmetry of these solutions is broken in a very ``weak" manner (hence the solutions are also called almost-BPS solutions). 
It is clear what this weak supersymmetry breaking means from the perspective of  D-branes: the almost-BPS solutions describe the multicenter generalization of the D2-D2-D2-$\overline{\rm D6}$ system (which is T-dual to the D4--D4--D4--$\overline{\rm D0}$ system), whose supersymmetry is broken because one of the component branes is not compatible to the supersymmetries preserved by the other three, and upon removing any one of the four charges supersymmetry is restored. 
However, from a supergravity perspective, characterizing solutions with controllable supersymmetry breaking is no easy task and these new ideas have been the cornerstone for subsequent developments. 
An interesting independent approach to extremal non-BPS solutions has been taken in \cite{Trigiante,Bossard,Kim:2010bf}, where the authors, making use of the symmetries of the solutions, reduce the problem to the study of a $\sigma$-model in one dimension less. 
Then, using the classification of nilpotent orbits, they are able to find non-BPS solutions, and to understand how they relate to the BPS ones (at least when the $\sigma$-model is given by a homogeneous manifold). 

In \cite{Bena:2009fi}, two of the authors with Bena and Warner have proposed a supergravity criterion for constructing non-BPS solutions where the equations of motion factorize: the existence of a ``floating brane'', or in more formal terms, the existence of a calibration. Upon imposing that in the supergravity ansatz a probe M2 brane feels no force, the supergravity equations of motion indeed factorize. The most general solution is constructed starting from an Israel--Wilson base space, and solving a set of linear equations for the fluxes and for the warp factor \cite{Bena:2009fi}. The solutions with an Israel--Wilson base are more general than both the BPS and the almost-BPS solutions, and reduce to them in certain limits. Hence, in the presence of calibrations, the equations of motion factorize, and this allows one to construct non-BPS solutions with relative ease. This phenomenon has also been observed in the construction of non-BPS flux compactifications \cite{Lust:2008zd,Cassani:2009na}.

While the almost-BPS solutions and solutions with an Israel--Wilson base that have been explicitly constructed \cite{Bena:2009ev,Bena:2009en,Bena:2009fi} comprise a large family of physically interesting geometries, and they allow one to understand, for example, the decay of multicenter non-supersymmetric solutions across lines of marginal stability in the moduli space, these solutions are not the most general multicenter extremal solutions. 
For instance, a group-theoretical analysis of the possible horizon configurations for two-centre solutions \cite{Ferrara:2010ru,Ferrara:2010cw,Ferrara:2010ug} shows an interesting landscape of stability regions and charge configurations that are not fully covered by the solutions in \cite{Bena:2009fi,Bena:2009ev,Bena:2009en} and for which explicit solutions should be constructed in order to put the analysis of \cite{Ferrara:2010ru,Ferrara:2010cw,Ferrara:2010ug} on firm grounds. Other hints for the existence of new non-BPS extremal solutions were also given in \cite{Kim:2010bf}. 

One route to enlarge the class of multicenter extremal solutions and to understand the structure underlying it is to perform a so-called spectral flow transformation \cite{Bena:2008wt}, which rotates some of the charges into others. 
This transformation leaves the class of BPS solutions invariant, but transforms for example almost-BPS solutions in Israel--Wilson solutions. 
However, writing this transformation explicitly is cumbersome, to say the least. 
A more straightforward, though technically challenging route, is to take a known nonextremal solution in type IIA supergravity compactified on a six-torus, and perform six T-dualities on it. 
This transforms, for example, the rotating D2-D2-D2-$\overline{\rm D6}$ solution found in \cite{Bena:2009ev} into a rotating D4-D4-D4-$\overline{\rm D0}$ black hole, which uplifts to the five-dimensional non-BPS extremal rotating M5-M5-M5-P of \cite{Kim:2010bf}. 
From a 4-dimensional perspective, both spectral flow transformations and T-dualities are just part of the same U-duality group of transformations that map the equations of motion and Bianchi identities into each other and hence map solutions into new solutions of the same system of equations.
One way to generate the most general multicenter extremal solution would be to first identify the 4-dimensional quantities related to a specific ``seed'' solution and then use an appropriate U-duality transformation to construct new configurations with arbitrary charges.

The purpose of this paper is to show explicitly how these duality transformations work, both from the perspective of eleven-dimensional supergravity compactified on $T^6$, and from the perspective of four-dimensional supergravity. Moreover, we concentrate on extremal multicenter non-BPS geometries of the under-rotating type, i.e. geometries that, like the extremal Reisner-Nordstr\"om black hole, have a conformally flat three-dimensional base. 
These solutions are different from the extremal Kerr-type solutions, in which the three spatial non-compact directions are described by a non-conformally flat metric. Within this under-rotating class, we obtain the most general explicit extremal multicenter solution of the $STU$ model.

As a first step, we check the action of the duality group on the BPS solutions.
From the ten-dimensional point of view, we show that starting from a BPS solution determined by 8 harmonic functions and performing 6 T-dualities, one obtains a BPS solution whose harmonic functions are simply interchanged. 
From the four-dimensional point of view, we also show that this is just a special case of the most general U-duality action, which rotates the harmonic functions by a symplectic matrix.

We then display the action of the duality transformations on almost BPS solutions. 
One of the features of these solutions is that one can turn on arbitrary Wilson lines at infinity (corresponding to axion vacuum expectation values in four dimensions) without changing the warp factors or the field strengths. However, when performing six T-dualities, two solutions that differ only in the axion vev at infinity are transformed into drastically-different solutions: the duality transformation that relates these two solutions is nothing but the spectral flow transformation of \cite{Bena:2008wt}. It should be recalled that, despite the fact that they have a 
clear CFT dual interpretation, spectral flow transformations are from a supergravity perspective far from simple, and the way they are encoded inside the U-duality group was everything but clear. The relation we find between spectral flows, T-duality and large gauge transformations is therefore highly non trivial, and helps us understanding the status of the spectral flows inside the U-duality group. Once again, this structure becomes more clear from a four-dimensional perspective, where all these transformations are part of the same duality group. We go on to compute the general U-duality action on the almost BPS solutions for the $STU$ model and show that for a specific choice of the duality parameters one recovers the solution generated by applying six T-dualities, spectral flow transformations and axion shifts to the ten-dimensional solutions.

We then see that the general class of extremal non-BPS multicenter under-rotating solutions presents a surprisingly rich structure: T-dualities and axion shifts can be used to relate different sub-classes of solutions and to generate, starting from the known almost-BPS solutions, more general multicenter solution for the $STU$ model. This solution is obtained by solving essentially the same equations as for almost-BPS solutions, except that the warp factors and electric fields are now complicated square roots of quadratic polynomials of the functions satisfying the almost-BPS equations. Our solution generating technique has useful physical applications, both for understanding the physics of multicenter extremal solutions and for constructing smooth microstates geometries for under-rotating extremal black holes in four and five dimensions. We leave the analysis of the physics of black hole and microstate solutions that can be constructed by our methods to later works.

In Section 2 we review BPS and non-BPS seed solutions in terms of eleven-dimensional supergravity. For our purposes, we reduce them first to solutions of type IIA supergravity, and then to solutions of the $STU$ model in four-dimensional N=2 supergravity. In Section 3 we work out, for our ansatz, the general field expressions after 6 T-dualities along each of the internal $T^6$ directions. We then apply these tranformations first to BPS solutions in section 4 and second to non-BPS ones in section 5. For the BPS case, we also find a relation between spectral flow transformations, T-dualities and axion shifts. This relation being more involved for the non-BPS solutions, we study it independently in section 6. Section 7 is devoted to some interesting subcases. In particular, we find a new rotating black string, and recover the class of solution based on an Israel-Wilson space. Finally, we generalized in section 8 the dualities performed before to the full U-duality group. This is done using a four-dimensional point of view. Technical details of the dimensional reduction and dualization of the gauge fields are given respectively in appendix A and B. Appendix C and D give details of the subcases studied in section 7. 

\vspace{.2cm}

\noindent
\emph{Note added:} Just before this paper was submitted, we received reference \cite{Galli:2010mg}, which derives a new set of first order equations for non-BPS multi-center black holes in 4-dimensional supergravity. We expect that a detailed comparison with the solutions presented here could shed light on how to construct the most general solution for an arbitrary scalar manifold.

\section{The setup}\label{sec:thesetup}

\subsection{The 11-dimensional ansatz}

The solutions we consider in the following are both supersymmetric and non-supersymmetric configurations of 11-dimensional supergravity carrying various M2, M5 and KK6 monopole and momentum charges.
We start by assuming a compactification to 5 dimensions by employing $(T^2)^3 \simeq T^6/Z_2 \times Z_2$ as internal space.
This is a simple prototype of compactifications on Calabi--Yau manifolds, which are natural extensions of our work.
We also fix the volume of the internal space to 1, as well as all the complex structure deformations.
The resulting ansatz for the metric and 3-form is
\begin{equation}
\label{11Dfields}
\begin{array}{rcl}
ds_{11}^2&=&\displaystyle -Z^{-2}(dt+k)^2+Z ds^2_4+\sum_{I=1}^3{Z\over Z_I} ds^2_I\,, \\[8mm]
A^{(3)}&=&\displaystyle\sum_{I=1}^3 A^{(3)}_I\wedge dT_I =\sum_{I=1}^3\left(-{dt+k\over Z_I}+a_I\right)\wedge dT_I\,,
\end{array}
\end{equation}
where the warp factor is fixed to $Z=(Z_1 Z_2 Z_3)^{1/3}$ and where $ds^2_I$ and $dT_I$ are the metric and the volume form on the I-th 2-torus, respectively. 
In detail, $ds^2_I=dy^2_{I,1}+dy^2_{I,2}$ and $dT_I=dy_{I,1}\wedge dy_{I,2}$.
Since we are also interested in the possibility of further reducing these configurations to 4 dimensions, we further specialize the metric $ds_4^2$ to that of a Gibbons--Hawking space
\begin{equation}
\label{GHmetric}
ds^2_4 = V^{-1}(d\psi+A)^2+V ds^2_3(\vec x)\,,
\end{equation}
with
\begin{equation}
  \label{basesign}
  \star dA= \pm dV,
\end{equation}
where $\star$ denotes Hodge duality in the 3-dimensional flat space $ds^2_3(\vec x)$ and the sign specifies the orientation.
Different choices of orientation lead to different types of solution. 
In particular, the plus sign corresponds to BPS configurations, while the minus sign leads to non supersymmetric solutions.

Once the 4-dimensional base space has been specialized, we need to decompose the 1-forms $a_I$ and $k$ accordingly
\begin{equation}
\label{1forms}
a_I = P_I \,(d\psi+A) + w^I \,,  \qquad k = \mu\, (d\psi+A) + \omega \,,
\end{equation}
so that $w^I$ and $\omega$ are 1-forms on $ds^2_3(\vec x)$.
At this point, the equations of motion governing the solutions to the 11-dimensional supergravity theory reduce to differential conditions in terms of the coordinates $\vec x$ of the 3-dimensional base of the Gibbons--Hawking space.
They read
\begin{eqnarray}
  d \star d Z_I &=& \frac{C_{IJK}}{2}\, d \star d( V P_J P_K), \nonumber \\[2mm]
  \star d w^I &=& - d(V P_I),  \label{BPSeq} \\[2mm]
  \star d \omega &=& V d \mu - \mu \, dV - V \, Z_I d P_I, \nonumber
\end{eqnarray}
for the BPS case (here $C_{IJK} = |\epsilon_{IJK}|$), i.e.~for the choice of plus sign in (\ref{basesign}), and
\begin{eqnarray}
  d \star d Z_I &=& \frac{C_{IJK}}{2} V d \star d(P_J P_K), \nonumber \\[2mm]
  \star d w^I &=& P_I\, dV - V dP_I, \label{nonBPSeq}\\[2mm]
  \star d \omega &=& d(\mu V) - V\,  Z_I d P_I, \nonumber
\end{eqnarray}
in the non-BPS case, i.e.~minus sign in (\ref{basesign}).
Solutions to the latter set of equations are also called almost-BPS, following \cite{Goldstein:2008fq,Bena:2009ev}.

The BPS equations (\ref{BPSeq}) admit simple solutions in terms of 8 harmonic functions named $\{V, K_I, M, L_I\}$ or $\{H^\Lambda, H_{\Lambda}\}$ in the context of the 11-dimensional analyses of \cite{Bena:2004de} and in the context of 4-dimensional supergravity \cite{Denef:2000nb}, respectively.
The relations with the functions appearing in (\ref{BPSeq}) is
\begin{eqnarray}\label{ZIBPS}
	P_I &=& \frac{K_I}{V} \,, \\[2mm]
	Z_I &=& L_I + \frac{C_{IJK}}{2} \frac{K_J K_K}{V}, \\[2mm]
	\mu &=& M + \frac{L_I K_I}{2V} + \frac{C_{IJK}}{6} \frac{K_I K_J K_K }{V^2} \,,
\end{eqnarray}
and, furthermore,
\begin{equation}\label{harm}
	V = -\sqrt2 \,H^0, \quad K_I = -\sqrt{2} \, H^I, \quad L_I = \sqrt2\, H_I, \quad M = \frac{1}{\sqrt2}\, H_0.
\end{equation}
The almost BPS equations (\ref{nonBPSeq}), on the other hand, cannot be solved in general only in terms of harmonic forms \cite{Bena:2009ev}.
Also in this case, however, $V$ is harmonic and one can easily verify that so is $P_I = K_I$.

In order to understand the detail of the chain of duality transformations we are going to perform in the following, it is useful to rewrite the configurations discussed above in terms of type IIA supergravity in 10 dimensions, as well as in terms of $N=2$ supergravity in 4 dimensions.

\subsection{Fields in type IIA} \label{fieldsIIA}

The type IIA configurations can be obtained by a direct reduction along the $\psi$ coordinate, which is a U(1) isometry of our solutions.
This leads to the following system of relations for the metric (in the string frame), dilaton and form-field potentials and the functions and forms of the 3-dimensional base defined above:
\begin{eqnarray}
\label{BPS10D}
ds^2_{10} &=& -e^{2 U}(dt+\omega)^2+e^{-2U} ds^2_3 + \sum_{I=1}^3 {e^{-2U}\over V Z_I } ds^2_I\,, \nonumber \\[2mm]
e^{-2\Phi} &=& e^{6 U} \,V^3 Z^3 \,, \nonumber \\[2mm]
B^{(2)} &=& \sum_{I=1}^3 B^{(2)}_I dT_I=\sum_{I=1}^3\left(P_I-{\mu\over Z_I}\right) dT_I \,, \\[3mm]
C^{(1)} &=& A-{\mu V^2 e^{4U}}(dt+\omega)\,, \nonumber \\[2mm]
C^{(3)} &=& \sum_{I=1}^3C^{(3)}_I\wedge dT_I=\sum_{I=1}^3 \left[-{dt+\omega\over Z_I} + \left(P_I-{\mu\over Z_I}\right) A+w^I\right] \wedge dT_I \,, \nonumber
\end{eqnarray}
where we introduced the positive definite quantity
\begin{equation}
e^{-4U} = Z_1 Z_2 Z_3 V - \mu^2 V^2\,.
\end{equation}

The Neveu--Schwarz (NSNS) 3-form and the Ramond--Ramond (RR) field strengths follow by simple differentiations of the above potentials
\begin{equation}
H^{(3)}=\sum_{I=1}^3 H^{(3)}_I \wedge dT_I= \sum_{I=1}^3 dB^{(2)}_I\wedge dT_I \,,
\end{equation}
and
\begin{equation}
\label{fieldstrength}
F^{(2)}=d C^{(1)}\,,\qquad F^{(4)}= \sum_{I=1}^3 F^{(4)}_I \wedge dT_I= \sum_{I=1}^3 \left(dC_I^{(3)}-H^{(3)}_I\wedge C^{(1)}\right)\wedge dT_I\,.
\end{equation}
Since we are interested in performing a series of T-duality transformations on the solutions to the (\ref{BPSeq}) and (\ref{nonBPSeq}) systems of equations, we give here also the dual gauge fields $C^{(5)}$ and $C^{(7)}$, which are related to the dual field strengths \begin{eqnarray}
F^{(6)}&=&-*_{10} F^{(4)} = d C^{(5)}- H^{(3)}\wedge C^{(3)}\,,\\[2mm]\label{fieldstrengthbis}
F^{(8)}&=&*_{10} F^{(2)}= dC^{(7)}-H^{(3)}\wedge C^{(5)}
\end{eqnarray}
($*_{10}$ is the Hodge dual with respect to the 10-dimensional string metric).

The details of the computations are given in the Appendix B for the BPS case, but the non-BPS case follows exactly in the same way. 
Once again, the structure of the original ansatz dictates the explicit form of these tensor fields, which can be summarized as
\begin{equation}
	C^{(5)} = \sum_{J,K=1}^3 C^{(5)}_{JK} \wedge dT_J \wedge dT_K,
\end{equation}
where
\begin{eqnarray}
C^{(5)}_{JK} &=& {\mu \over Z_J Z_K} (dt + \omega) - C_{JKI}\, v_I + \left( P_J - {\mu\over Z_J} \right) \left( P_K - {\mu\over Z_K} \right) A \nonumber \\[2mm]
&+&  \left( P_J - {\mu\over Z_J} \right) w^K + \left( P_K - {\mu\over Z_K} \right) w^J\,,
\end{eqnarray}
and
\begin{eqnarray}
C^{(7)} &=& \left[ C^{(7)}_t (dt+\omega) -v_0 - \left( P_I - {\mu\over Z_I} \right) v_I + \left( P_1 - {\mu\over Z_1} \right) \left( P_2- {\mu\over Z_2} \right) \left( P_3 - {\mu\over Z_3} \right) A \right. \nonumber \\[2mm]
 && \left.\quad + {C_{IJK}\over 2} \left(P_I- {\mu\over Z_I} \right) \left( P_J - {\mu\over Z_J} \right) w^K  \right] \wedge dT_1 \wedge dT_2 \wedge dT_3 \,.
\end{eqnarray}
Although the general structure of these forms is the same in the BPS and non-BPS cases, the detailed expressions for the time component of $C^{(7)}$ and the 1-forms $v_0$ and $v_I$ are different in the two cases. These are given by
\begin{eqnarray}
\star d v_0 &=& 2\, d\left[\mu - \frac12 Z_I P_I +\frac{1}{2} V P_1P_2P_3 \right] = 2\,  dM\,,\\[3mm]
\star d v_I &=&  d Z_I - \frac{C_{IJK}}{2} d\left[V P_J P_K\right] = dL_I \,, 
\end{eqnarray}
\begin{equation}
  C^{(7)}_t ={e^{-4U}\over V^2 Z^3},
\end{equation}
%
for the BPS case and
\begin{eqnarray}
\star d v_0 &=&  Z_I d P_I - P_I dZ_I + V d (P_1P_2P_3) - (P_1P_2P_3) d V \,\label{defv0},\\[3mm]
\star d v_I &=& d Z_I - \frac{C_{IJK}}{2}\left[V d(P_J P_K) -P_J P_K dV\right] \,, 
\end{eqnarray}
\begin{equation}
  C^{(7)}_t ={e^{-4U}\over V^2 Z^3} - \frac2V,
\end{equation}
for the almost-BPS solutions.

\subsection{Fields in 4-dimensional $N=2$ supergravity}

The combined reduction on the $\psi$ direction and on the $T^6/Z_2 \times Z_2$ internal space gives a 4-dimensional configuration that can be expressed in the language of $N=2$ supergravity coupled to 3 vector multiplets, whose scalar manifold parameterizes the $STU$ model.
The bosonic lagrangian of the reduced theory is specified by few data that can be expressed in terms of the functions appearing in the ansatz employed at the beginning of this section.

The bosonic sector of Einstein--Maxwell $N=2$ supergravity is
\begin{equation}
	{\cal L}_{4d} = \frac12\, R - g_{i \bar \jmath}\partial_\mu z^i \partial^\mu \bar z^{\bar \jmath} + \frac18\, {\cal I}_{\Lambda \Sigma} F_{\mu\nu}^\Lambda F^{\Sigma\,\mu\nu} + \frac18\, {\cal R}_{\Lambda \Sigma} F_{\mu\nu}^\Lambda (*_4 F)^{\Sigma\,\mu\nu},
	 \label{StartingAction} 
\end{equation}
where $I$ runs over the number of vector multiplets (in our case $I=1,2,3$), $\Lambda = \{0,I\}$ includes also the index associated to the graviphoton, so that $F^\Lambda = d A^\Lambda$ and $*_4$ is the Hodge duality operation in 4 dimensions (which we defined via $\epsilon_{0123} = 1$).

The configurations we study in this paper are solutions to the equations of motion of this system with a specific ansatz for the various fields.
We give the details on the identification procedure in appendix A, but we quote the results here.
The 4-dimensional metric is given by
\begin{equation}
  ds^2_{4d} = - {\rm e}^{2U} (dt + \omega)^2 + {\rm e}^{-2 U} ds_3^2(\vec{x}),
\end{equation}
which is the appropriate form for studying stationary solutions including generic multi-center as well as extremal under-rotating single center black hole configurations.
The 11-dimensional ansatz used to derive both BPS and almost-BPS solutions implies a constrained form for the three scalar fields, which read
\begin{equation}
  z^I = \frac{\left(V Z_I P_I -  V \mu\right) - i\, {\rm e}^{-2 U}}{V Z_I} \label{zI}
\end{equation}
and for the vector fields $A^\Lambda$, which are 
\begin{eqnarray}
	A^0 &=& w^0 + {\rm e}^{4U} \,\mu\, V^2 (dt +  \omega), \qquad w^0 = -A,\label{vec0} \\[4mm]
	A^I &=& w^I - \frac{e^{4U}V}{Z_I} \left(Z_1 Z_2 Z_3-\mu V P_I Z_I\right)(dt + \omega). \label{vecI}
\end{eqnarray}
The remaining couplings of the 4-dimensional theory are functions of the scalar fields and are constrained by the geometry of the scalar manifold for the $STU$ model, namely 
\begin{equation}
  {\cal M}_{scalar} = \left[\frac{\rm SU(1,1)}{\rm U(1)}\right]^3.
\end{equation}
The metric of the scalar $\sigma$-model $g_{I\bar J}$ follows from the K\"ahler potential
\begin{equation}
  K = - \log(-i (z^1-\bar z^1)(z^2-\bar z^2)(z^3-\bar z^3))
\end{equation}
and the gauge kinetic couplings ${\cal I}$ and ${\cal R}$ are detailed in the appendix A.

Also in 4 dimensions it is useful to introduce the dual field strengths $G_{\Lambda}$.
These are defined as
\begin{equation}
	\label{Hodgeduality}
 	G_\Lambda = {\cal R}_{\Lambda \Sigma} F^\Sigma - {\cal I}_{\Lambda \Sigma} *_4 F^\Sigma ,
\end{equation}
so that the Bianchi identities of these field strengths coincide with the equations of motion of the original vector fields $A^\Lambda$.
Since the equations of motion of the vector fields in (\ref{StartingAction}) are $d G_{\Lambda} =0$, we can introduce (locally) dual vector potentials $A_{\Lambda}$, so that $G_{\Lambda} = d A_{\Lambda}$.
Given the form of the ansatz presented previously and the form of the 4-dimensional metric, we can split the electric $A^\Lambda$ and magnetic $A_{\Lambda}$ vector fields as
\begin{eqnarray}
	\label{ALup}
	A^\Lambda &=& w^\Lambda + \chi^\Lambda (dt + \omega), \\
	A_\Lambda &=& v_\Lambda + \psi_\Lambda (dt + \omega).\label{ALdown}
\end{eqnarray}
Using these expressions in the relation (\ref{Hodgeduality}) we can obtain an explicit procedure to compute the dual potentials.
First we use the electric vector fields components to obtain $\psi_{\Lambda}$
\begin{equation}
  \label{dpsi}
	d \psi_\Lambda = {\rm e}^{2 U}{\cal I}_{\Lambda \Sigma} \star \left(d w^\Sigma+ \chi^\Sigma d \omega\right) + {\cal R}_{\Lambda \Sigma} d \chi^\Sigma
\end{equation}
(consistency imposes $d^2 \psi_{\Lambda} = 0$), and then plug the solution into 
\begin{equation}
  \label{dAdual}
	\star d v_{\Lambda} = -{\rm e}^{-2 U}{\cal I}_{\Lambda \Sigma} \, d \chi^\Sigma +{\cal R}_{\Lambda \Sigma} \star \left(d w^\Sigma + \chi^\Sigma d \omega\right) - \psi_{\Lambda} \star d \omega.
\end{equation}
to obtain the expression for $v_{\Lambda}$.

\subsubsection{The BPS case}

As explained above, in the BPS case, the solutions can be expressed in terms of eight harmonic functions as in (\ref{harm}).
This simplifies further the expression of the various 4-dimensional quantities.
The warp factor reduces to
\begin{equation}
\begin{array}{rcl}
e^{-4 U} &=& \displaystyle V L_1 L_2 L_3 - 2M K_1 K_2 K_3 - M^2 V^2 - M V \sum_I L_I K_I \\[2mm]
&+&\displaystyle {1\over 2} \sum_{I<J} L_I K_I L_J K_J - {1\over 4} \sum_I L_I^2 K_I^2 = I_4(H^\Lambda, H_{\Lambda}) \,,
\end{array}
\end{equation}
where
\begin{equation}
	I_4(p^\Lambda,q_{\Lambda}) = -\left(p^\Lambda q_{\Lambda}\right)^2 + 4 \sum_{I< J}(p^I q_I p^J q_J)-4 p^0 q_1 q_2 q_3 + 4 q_0 p^1 p^2 p^3
	\label{I4def}
\end{equation}
is the so-called quartic invariant of the $STU$ model.
We point out that the sign of the last term depends on the definitions of the invariant and usually it is taken with a minus sign in papers dealing with the 10-dimensional constructions \cite{Bena:2005va}, while it is taken with a plus in supergravity literature \cite{Behrndt:1996hu}.
Also the scalar fields (\ref{zI}) can be expressed in terms of the harmonic functions by
\begin{equation}
	z^I = \frac{H^I + i\, \frac{\displaystyle \partial \sqrt{I_4}}{\displaystyle\partial H_I}}{H^0 + i\, \frac{\displaystyle\partial \sqrt{I_4}}{\displaystyle\partial H_0} },
	\label{scalarBPS}
\end{equation}
which agrees with the known solution for the scalar fields in a BPS solution \cite{Kallosh:2006ib}.
The 4-dimensional gauge potentials can also be identified by plugging in (\ref{vec0}) and (\ref{vecI}) the solutions (\ref{ZIBPS})--(\ref{harm}).
This implies 
\begin{eqnarray}
	\chi^0 &=& \sqrt2\, e^{4U} \left[H^0 H^\Lambda H_{\Lambda} -2 H^1 H^2 H^3\right], \label{chi0} \\[2mm]
	\chi^I &=& \sqrt2\, e^{4U} \left[H^I\left( H^\Lambda H_{\Lambda}  -2 \sum_{J \neq I} H^J H_J \right)+C_{IJK} H^0 H_J H_K\right],\label{psiI}
\end{eqnarray}
as well as
\begin{equation}
	\star dw^\Lambda = \sqrt2\, dH^\Lambda.
\end{equation}

The dual gauge potentials are obtained by solving (\ref{dpsi}):
\begin{eqnarray}
	\psi_{0} &=& -\sqrt2 \, e^{4U}  \left[H_0 H^\Lambda H_{\Lambda}+2 H_1 H_2 H_3\right], \\[2mm]
	\psi_I &=& \sqrt2 \, e^{4U} \left[- H_I  \left(H^\Lambda H_{\Lambda} - 2 \sum_{J \neq I}H^J H_J\right) + C_{IJK} H_0 H^J H^K)\right].
\end{eqnarray}
We then use this solution (\ref{dAdual}) to determine the expression for $v_{\Lambda}$, which, again in terms of the harmonic functions, reaches the simple form
\begin{equation}
	\star dv_{\Lambda} = \sqrt{2}\, d H_{\Lambda}. \label{stardv}
\end{equation}
The factors of $\sqrt2$ appearing in front of the harmonic functions are related to the normalization of the vector fields used in the lagrangian (\ref{StartingAction}), which follows from the reduction procedure presenting in the appendix.
A canonical normalization of the vector field terms in (\ref{StartingAction}) by a factor of $1/4$ rather than $1/8$ would get rid of the square roots.
However, we preferred to keep these factors, so that we can identify the 4-dimensional vector fields directly with the related components of the $C^{(3)}$, $C^{(5)}$ and $C^{(7)}$ form fields.

\subsubsection{The almost-BPS case} \label{sub:almostBPS}

The generic non-BPS solution cannot be expressed entirely in terms of harmonic functions, but in the special case of single centre non supersymmetric black holes.
However, we can associate the functions $P_I$ to harmonic functions ($K_I$ in \cite{Bena:2009ev}).
The initial data are then 
\begin{equation}\label{chisingle}
	\begin{array}{lcl}
	\chi^0 = e^{4U}{\mu V^2}, && \displaystyle \chi^I = e^{4U}V \left(\mu V K_I - \frac12 C_{IJK} Z_J Z_K\right),\\[6mm]
	 \star d w^0 = dV, && \star d w^I = K_I dV - V d K_I.
	\end{array}
\end{equation}
Once again, from the definition of the Hodge dual gauge potentials (\ref{dpsi}), we get 
\begin{equation}
	\label{psisingle}
	\begin{array}{rcl}
	\psi_0 &=&  e^{4U}\left[Z_1 Z_2 Z_3 - \mu V\left(V K_1 K_2 K_3 + \sum_I Z_I K_I\right)+ V \sum_{J< K} (K_J K_K Z_J Z_K)\right], \\[4mm]
	\psi_I &=&  e^{4U} V \left[Z_I\left(\mu- \sum_{J \neq I}K_J Z_J\right)+\frac{C_{IJK}}{2}V \mu K_J K_K\right] 
\end{array}
\end{equation}
and, plugging this result into (\ref{dAdual}), 
\begin{eqnarray}
	\star dv_0 &=&  Z_I d K_I - K_I dZ_I+V d(K_1 K_2 K_3)- (K_1 K_2 K_3) dV , \\[2mm]
	\star dv_I &=& d Z_I - \frac{C_{IJK}}{2}[V d(K_J K_K) - K_J K_K dV]. 
\end{eqnarray}

In the single center case, one can express the most general solution in terms of 4 harmonic solutions \cite{Lopes Cardoso:2007ky,Gimon:2007mh,Bellucci:2008sv}.
The generic solution in this case can be obtained by acting with the duality group on a so-called  seed solution, which falls in our ansatz whenever it is constituted by D2 and D6 charges.
In this instance the non-BPS equations (\ref{nonBPSeq}) are solved by setting $P_I = 0$ and by introducing a Taub-NUT charge in 
\begin{equation}
	V = \sqrt{2}\, H^0,
\end{equation}
which is given by a harmonic function, and electric charges in
\begin{equation}
	Z_I = \sqrt{2}\, H_I,
\end{equation}
which also become harmonic.
Finally a non-trivial axion can be turned on by imposing
\begin{equation}
	\mu = \frac{b}{V}.
\end{equation}
This means that
\begin{equation}
	e^{-4U} = 4 H^0 H_1 H_2 H_3 - b^2,
\end{equation}
which is now identified with $-I_4$ when $b=0$ and we approach the horizon of the black hole solution.
This is consistent with the fact that the quartic invariant changes sign between the BPS and non-BPS single centre solutions.

\section{T-dualities and Buscher's rules}\label{sec:Tdualities} 

Starting from the BPS and the non-BPS configurations described in section \ref{sec:thesetup}, we will now apply various duality transformations to generate new solutions.
In particular, we will focus on the action of T-duality along all the directions of $T^6$: $z_{1,1}, z_{1,2},\ldots z_{3,1},z_{3,2}$. 
This part of the computation equally applies to the BPS and non-BPS case. 

As it is known \cite{Bergshoeff:1995as,Hassan:1999bv}, T-duality transformations act on the supergravity fields mixing them according to Buscher's rules, which we now summarize in order to fix our conventions.
If we assume that $y$ is the direction along which one performs the T-duality trasformation and that the string metric, B-fields and RR gauge fields $C^{(p)}$ split according to
\begin{eqnarray}
ds_{10}^2 &=& G_{yy} (dy+A_\mu dx^\mu)^2 + \widehat{g}_{\mu\nu} dx^\mu dx^\nu\,,\nonumber \\[2mm]
B^{(2)}&=& B_{\mu y} dx^\mu \wedge (dy+A_\mu dx^\mu) + \widehat{B}^{(2)}\,, \label{trule0}\\[2mm]
C^{(p)}&=&C_y^{(p-1)}\wedge  (dy+A_\mu dx^\mu) + \widehat{C}^{(p)}\,,\nonumber
\end{eqnarray}
where the forms $\widehat{B}^{(2)}$, $C_y^{(p-1)}$ and $\widehat{C}^{(p)}$ do not have legs along $y$ and are functions only of the $x^\mu$ coordinates, the T-duality transformed fields are
\begin{eqnarray}
d{\widetilde s}_{10}^2&=& G^{-1}_{yy} (dy-B_{\mu y} dx^\mu)^2 + \widehat{g}_{\mu\nu} dx^\mu dx^\nu\,,\quad e^{2 \widetilde \Phi}={e^{2\Phi}\over G_{yy}}\,,\nonumber \\[2mm]
{\widetilde B}^{(2)}&=& -A_{\mu} dx^\mu dy + \widehat{B}^{(2)}\,, \label{trule}\\[2mm]
{\widetilde C}^{(p)}&=& \widehat{C}^{(p-1)}\wedge(dy-B_{\mu y} dx^\mu)+C_y^{(p)}\,.\nonumber
\end{eqnarray}
Equivalently, the rules on the RR forms can be written as
\begin{equation}
\begin{array}{l}
\displaystyle\widetilde{C}^{(n)}_{\mu...\nu\alpha y} = C^{(n-1)}_{\mu...\nu\alpha}
- (n-1)\frac{C^{(n-1)}_{[\mu...\nu | y} g_{|\alpha]y}}{g_{yy}}\,, \\\\
\displaystyle\widetilde{C}^{(n)}_{\mu...\nu\alpha\beta} =
C^{(n+1)}_{\mu...\nu\alpha\beta y} + n
C^{(n-1)}_{[\mu...\nu\alpha}B_{\beta]y} +
n(n-1)\frac{C^{(n-1)}_{[\mu...\nu | y}B_{|\alpha|y}
g_{|\beta]y}}{g_{yy}} \,.
\end{array}
\end{equation}

In the case at hand, the IIA fields we are dealing with have a special form that simplify the expression of the dual fields if we apply a sequence of couples of T-dualities along the directions of the same two-torus.
In fact, the structure of the metric and form fields is
\begin{eqnarray}
&& ds^2_{10} = ds^2_{4d} + \sum_I G_I ds^2_I \,, \quad B^{(2)} = \sum_I B_I dT_I \,, \\[2mm]
&& C^{(3)} = \sum_I C^{(3)}_I \wedge dT_I \,, \quad C^{(5)} = \sum_{I<J} C^{(5)}_{IJ} \wedge dT_I \wedge dT_J \,, \quad C^{(7)} = C^{(7)} \wedge dT_1 \wedge dT_2 \wedge dT_3 \,, \nonumber
\end{eqnarray}
and the sequence of two T-dualities along $z_{I,1}, z_{I,2}$ on the NSNS fields can be seen as a simple inversion of the matrix 
\begin{equation}
  E_I= \begin{pmatrix} G_I & B_I \\ -B_I & G_I \end{pmatrix} \,,
\end{equation}
which means
\begin{equation}
\label{NSNStsfo}
E_I\longrightarrow \widetilde E_I= E^{-1}_I = {1\over \Delta_I} \begin{pmatrix} G_I&-B_I\\ B_I& G_I\end{pmatrix}\,,\quad \Delta_I=G_I^2+B_I^2\,,
\end{equation} 
and a rescaling of the dilaton
\begin{equation}
e^{2\Phi}\longrightarrow e^{2\widetilde \Phi}= {e^{2\Phi}\over \Delta_I}\,.
\end{equation}
At the same time, the RR fields transform as
\begin{eqnarray}
&& \widetilde C^{(1)}=-C^{(3)}_I + B_I C^{(1)}\,,\\[2mm]
&& \widetilde C^{(3)}_I =\Delta_I^{-1} (B_I C^{(3)}_I + G_I^2 C^{(1)})\,,\quad \widetilde C^{(3)}_J = -C^{(5)}_{JI}+B_I C^{(3)}_J \,\,(J\not=I)\,, \\[2mm]
&& \widetilde C^{(5)}_{IJ}= \Delta_I^{-1}(B_I C^{(5)}_{IJ}+G_I^2 C^{(3)}_J)\,\,(J\not =I)\,, \\[2mm]
&& \widetilde C^{(5)}_{JK}=-C^{(7)}+B_I C^{(5)}_{JK}\,\,(J\not=K,\,J,K\not=I)\,, \\[2mm]
&& \widetilde C^{(7)}=\Delta_I^{-1} (B_I C^{(7)}+G_I^2 {C_{IJK}\over 2}C^{(5)}_{JK})\,.
\end{eqnarray}
Iterating these rules on the three 2-tori, one finds the fields after 6 T-dualities:
\begin{eqnarray}
\label{6Ttsfo}
&&\!\!\!\!\!\!d\widetilde s^2_{10} = ds^2_{4d} +\sum_I {G_I\over \Delta_I} ds^2_I \,,\quad \widetilde B^{(2)}=-\sum_I{ B_I\over \Delta_I} dT_I \,,\\
&&\!\!\!\!\!\!  e^{2\widetilde \Phi}= {e^{2\Phi}\over \Delta_1 \Delta_2 \Delta_3}\,,\\
&&\!\!\!\!\!\!\widetilde C^{(1)}=-C^{(7)}+{C_{IJK}\over 2}B_I C^{(5)}_{JK}-{C_{IJK}\over 2} B_I B_J C^{(3)}_K + B_1 B_2 B_3 C^{(1)}\,,\\
&&\!\!\!\!\!\!\widetilde C^{(3)}_I = \Delta_I^{-1}\Bigl[B_I C^{(7)} -B_I C_{IJK} B_J C^{(5)}_{IK}+G_I^2 {C_{IJK}\over 2}C^{(5)}_{JK}\nonumber\\
&&\!\!\!\!\!\!\qquad\quad +B_1 B_2 B_3 C^{(3)}_I - G_I^2 C_{IJK} B_J C^{(3)}_K + G_I^2 {C_{IJK}\over 2} B_J B_K C^{(1)}\Bigr]\,,\\
&&\!\!\!\!\!\!\widetilde C^{(5)}_{IJ}=(\Delta_I\Delta_J)^{-1} \Bigl[-B_I B_J C^{(7)} + B_1 B_2 B_3 C^{(5)}_{IJ} -C_{IJK} (G_I^2 B_J C^{(5)}_{JK}+G_J^2 B_I C^{(5)}_{IK})\nonumber\\
&&\!\!\!\!\!\!\qquad\quad +C_{IJK} B_K (G_I^2 B_J C^{(3)}_J+ G_J^2 B_I C^{(3)}_I)-G_I^2 G_J^2 C_{IJK} C^{(3)}_K + G_I^2 G_J^2 C_{IJK} B_K C^{(1)} \Bigr]\,,\\
&&\!\!\!\!\!\!\widetilde C^{(7)}=(\Delta_I\Delta_J\Delta_K)^{-1}\!\Bigl[B_1 B_2 B_3 C^{(7)}+ {C_{IJK}\over 2} (G_I^2 B_J B_K C^{(5)}_{JK}+ G_I^2 G_J^2 B_K C^{(3)}_K)+ G_1^2 G_2^2 G_3^2 C^{(1)}\Bigr]\,.\nonumber
\end{eqnarray}
%

\section{Dualities and the BPS solutions}\label{sec:BPSdualities}

Having set the stage in detail in the previous sections, we can now apply the duality relations to the BPS configurations, solutions of (\ref{BPSeq}).
Comparison of the results of section \ref{fieldsIIA} with those of section \ref{sec:Tdualities}, we identify the IIA metric, dilaton and B-field with
\begin{eqnarray}
	G_I &=& {e^{-2U}\over Z_I V}= {e^{-2U}\over L_I V +{C_{IJK}\over 2} K_J K_K }\,, \\[2mm]
	B_I &=& {K_I\over V}-{\mu\over Z_I}= {-2 M V + 2 L_I K_I -\sum_A L_A K_A  \over 2 (L_I V +{C_{IJK}\over 2}  K_J K_K )}\,,\\[2mm]
	e^{-2\Phi} &=& {e^{6U} Z^3 V^3}= {e^{6U} \prod_{I} (L_I V +{C_{IJK}\over 2} K_J K_K )}\,
\end{eqnarray}
and, similarly, the fields obtained after 6 T-dualities on the directions of the ${T}^6$ are
\begin{eqnarray}
\widetilde G_I &=& {e^{-2U}\over -2 M K_I+{C_{IJK}\over 2} L_J L_K}\,, \\[2mm]
\widetilde B_I &=& {-2 M V + 2 L_I K_I -\sum_A L_A K_A  \over 2( -2 M K_I+{C_{IJK}\over 2} L_J L_K)}\,,\\[2mm]
e^{-2\widetilde \Phi}&=& {e^{6U} \prod_I (-2 M K_I+{C_{IJK}\over 2} L_J L_K)}\,.
\end{eqnarray}
A simple comparison of the two sets of expressions above shows that the series of 6 T-duality transformations on $T^6$ can be summarized by an exchange of the harmonic functions
\begin{equation}
 \widetilde{V} = 2M \,, \quad \widetilde{K}_I = L_I \,, \quad \widetilde{L}_I = - K_I \,, \quad \widetilde{M} = - {V \over 2} \,,
\label{bpst6}
\end{equation}
or
\begin{equation}
	\widetilde H^\Lambda = - H_\Lambda, \qquad \widetilde H_\Lambda = H^\Lambda.
\end{equation}
These transformations do not change the quartic invariant $I_4$, which explains its name.
Actually, the $I_4$ combination is actually a full U-duality invariant, as we will see later on.

For the RR fields, using again the explicit expressions of the fields in terms of harmonic functions appropriate for the BPS case, one can verify that the transformed RR fields $\widetilde C^{(p)}$ are related to the starting ones $C^{(p)}$ by the transformation (\ref{bpst6}), together with the corresponding transformation on the 1-forms
\begin{equation}
\widetilde{A} = v_0 \,,\quad \widetilde{w}_I = -v_I \,,\quad \widetilde{v}_0 = -{A} \,,\quad \widetilde{v}_I = w^I\,.
\label{bpst62}
\end{equation}
We thus reach the conclusion that T-duality on $T^6$ is equivalent, in the BPS case, to the transformations (\ref{bpst6})-(\ref{bpst62}).

\subsection{The BPS black string}\label{bpsstring:sec}

Since we are going to discuss a non-BPS black string in section \ref{KKK=0}, we will now quickly review how one can obtain a BPS black string. 

As we have seen above, performing 6 T-dualities exchanges $V$ with $M$ and $K_I$'s with $L_I$'s. The physical meaning of this operation is to exchange D0-branes with D6-branes and D2-branes with D4-branes.  One can therefore obtain a BPS black string, which has D4-D4-D4-D0 charges, by performing 6 T-dualities on a D6-D2-D2-D2 black hole. 

One might also ask if the  D4-D4-D4-D0 black string solution is already contained in the original (i.e. before T-duality) BPS ansatz (\ref{11Dfields}-\ref{BPSeq}), which in general carries also D6 and D2 charges. At first sight this might seem impossible, because the general BPS solution preserves the supersymmetries corresponding to the D6-D2-D2-D2 system, and not those of the D0-D4-D4-D4 system. To obtain a black string one should take the limit $V=0$ and $L_I=0$ in the general BPS solution, and in this limit the quantities describing the solution -- $Z_I$, $a_I$, $\mu$ -- diverge. It turns out however that these quantities combine in such a way as to give a finite limit 
for the full physical metric: for example, though $Z=(Z_1 Z_2 Z_3)^{1/3}$ diverges as $V^{-1}$, the 3-dimensional part of the metric is proportional to $Z V$ and is thus finite in the limit.

One then obtains the following solution
\begin{eqnarray}
 ds_{11}^2 &=& - \frac{2}{K} dt d\psi - \frac{2M}{K} d\psi^2 + K^2 ds_3^2 + \sum_I \frac{K_I}{K} ds_I^2 \,, \\ \nonumber
 A^{(3)} &=& \sum_I w^I \wedge dT_I \,,
\end{eqnarray}
where the quantity $K = (K_1 K_2 K_3)^{1/3}$ has been introduced. Taking
\begin{equation}
-2 M = 1+ {Q^{(D0)}\over r}\,,\quad K_I = 1 + {Q^{(D4)}_I\over r}
\end{equation}
this solution reproduces the well known black string solution, although in a system of coordinates 
which is not explicitly asymptotically flat. To rewrite the metric in a frame which is explicitly flat at asymptotic infinity one has to perform the coordinate redefinition
\begin{eqnarray}
\psi'=\psi -t\,,
\end{eqnarray}
and it is this change of coordinates which effectively transforms the supersymmetries preserved by the solution from those associated to D6-D2-D2-D2 to those of D0-D4-D4-D4.

\subsection{Gauge transformations, T-dualities and spectral flow} \label{gaugeSFBPS}

We have discussed until now only a small subset of the duality group of the system, given by T-dualities along the directions of $T^6$. Another important set of transformations that can be used to generate new solutions is given by spectral flows \cite{Bena:2008wt}. A spectral flow transformation is the composition of a U-duality transformation from the M2-M2-M2 to the D1-D5-P frame, where the D1 and D5 branes share the direction $y$ of $T^6$, and the coordinate redefinition
\begin{eqnarray}
\psi \to \psi + \gamma\, y\,.
\end{eqnarray}
As there are three inequivalent choices for the direction $y$ inside $T^2\times T^2\times T^2$, there are three different spectral flow transformations, whose parameters we will denote as $\gamma_I$. The physical interest of this transformation stems from the fact that it is dual to a well-known symmetry of the D1-D5 CFT, and it can be used to relate 2-charge (D1-D5) solutions to 3-charge (D1-D5-P) solutions \cite{Balasubramanian:2000rt,Maldacena:2000dr, Lunin:2004uu, Giusto:2004id, Ford:2006yb, AlAlawi:2009qe}. 

We will show in this subsection that spectral flows are equivalent to a combination of T-dualities and large gauge transformations of the type IIA B-field.

Consider a large gauge transformation that shifts the asymptotic value of the IIA B-field
\begin{equation}
	B^{(2)} \to 	B^{(2)} - \sum_{I=1}^3\gamma_I\, dT_I\,,
\end{equation}
and leaves the other fields invariants. This transformation is equivalent to the following redefinitions of the harmonic functions:
\begin{eqnarray}
\label{ggetsfoBPS}
 \widehat{V} &=& V \,, \quad \widehat{\omega} = \omega \,, \quad \widehat{K}_I = K_I -\gamma_I V \,, \nonumber \\
 \widehat{L}_I &=& L_I + C_{IJK} \gamma_J K_K - \frac{1}{2} C_{IJK} \gamma_J \gamma_K V \,, \\ \nonumber
 \widehat{M} &=& M + \frac{1}{2} \gamma_I L_I + \frac{1}{4} C_{IJK} \gamma_I \gamma_J K_K - \frac{1}{12} C_{IJK} \gamma_I \gamma_J \gamma_K V \,.
\end{eqnarray}
This operation maps solutions of (\ref{BPSeq}) to new solutions of the same set of equations.
Notice that this transformation acts trivially on solutions with four non-compact spatial directions:
since in that case the harmonic function $V$ vanishes at infinity, the transformation does not change the asymptotic value of the fields, and hence it reduces to a proper gauge transformation.
The situation is different in the case of interest for this paper, in which $\psi$ is a compact direction and correspondingly $V$ goes to a constant non-zero value at infinity. In this case solutions related by the transformation (\ref{ggetsfoBPS}) are physically inequivalent, as they have different 
values for the Wilson lines of the B-field.  Though this might seem a quite trivial difference, we will see that the large gauge transformation (\ref{ggetsfoBPS}) combined with T-dualities will give rise to more drastic effects, including different values for the asymptotic charges.

To understand this point, let us compare the two following solutions:\\
1) start from a solution encoded by the harmonic functions $(V,K_I,L_I,M)$ and perform on it
6 T-dualities, to arrive at the solution $(\widetilde{V}, \widetilde{K}_I,\widetilde{L}_I,\widetilde{M})$, given in eq. (\ref{bpst6});\\
2) on the solution $(V,K_I,L_I,M)$ perform first a large gauge transformation, giving the solution $(\widehat{V}, \widehat{K}_I,\widehat{L}_I,\widehat{M})$ in eq. (\ref{ggetsfoBPS}), and then 6 T-dualities; the resulting harmonic functions $(\widetilde{\widehat{V}},\widetilde{\widehat{K}}_I,\widetilde{\widehat{L}}_I,\widetilde{\widehat{M}})$ are given by:
\begin{eqnarray}
\widetilde{\widehat{V}} &=& 2 \widehat{M} = 2 M +\gamma_I L_I + \frac{1}{2} C_{IJK} \gamma_I \gamma_J K_K - \frac{1}{6} C_{IJK} \gamma_I \gamma_J \gamma_K V \,, \\ \nonumber
\widetilde{\widehat{K}}_I &=& \widehat{L}_I = L_I + C_{IJK} \gamma_J K_K - \frac{1}{2} C_{IJK} \gamma_J \gamma_K V \,, \nonumber\\ 
\widetilde{\widehat{L}}_I &=& -\widehat{K}_I = -K_I + \gamma_I V   \,, \nonumber \\
 \widetilde{\widehat{M}} &=& -{\widehat{V} \over 2} = -{V \over 2}\,.
\end{eqnarray}
Let us now compare the two solutions $(\widetilde{V}, \widetilde{K}_I,\widetilde{L}_I,\widetilde{M})$ and $(\widetilde{\widehat{V}},\widetilde{\widehat{K}}_I,\widetilde{\widehat{L}}_I,\widetilde{\widehat{M}})$:
\begin{eqnarray}
\label{spectralflow}
\widetilde{\widehat{V}} &=&  \widetilde{V} + \gamma_I \widetilde{K}_I - \frac{1}{2} C_{IJK} \gamma_I \gamma_J \widetilde{L}_K + \frac{1}{3} C_{IJK} \gamma_I \gamma_J \gamma_K \widetilde{M}  \,, \nonumber\\
\widetilde{\widehat{K}}_I &=& \widetilde{K}_I - C_{IJK} \gamma_J \widetilde{L}_K + C_{IJK} \gamma_J \gamma_K \widetilde{M} \,, \nonumber\\ 
\widetilde{\widehat{L}}_I &=& \widetilde{L}_I - 2 \gamma_I \widetilde{M} \,, \nonumber \\
 \widetilde{\widehat{M}} &=& \widetilde{M} \,.
\end{eqnarray}
The transformation above is exactly the spectral flow transformation as given in \cite{Bena:2008wt}. 
Since the T-duality map (\ref{bpst6}) is obviously invertible, we can equivalently state the result (\ref{spectralflow}) via the identity
\begin{eqnarray}
\label{spectralgaugelink}
\mathcal{F}(\gamma_I)=\mathcal{T}_6 \circ \mathcal{G}(\gamma_I) \circ \mathcal{T}_6^{-1}\,,
\end{eqnarray}
where $\mathcal{F}(\gamma_I)$ is the spectral flow operation, $\mathcal{G}(\gamma_I)$ is a large gauge transformation acting on the B-field, and $\mathcal{T}_6$ is the map resulting from 6 T-dualities on $T^6$. We have shown that the identity (\ref{spectralgaugelink}) holds on BPS solutions.

\begin{figure} \label{diag6TSF}
\centering
\includegraphics[angle=270,width=0.6\linewidth]{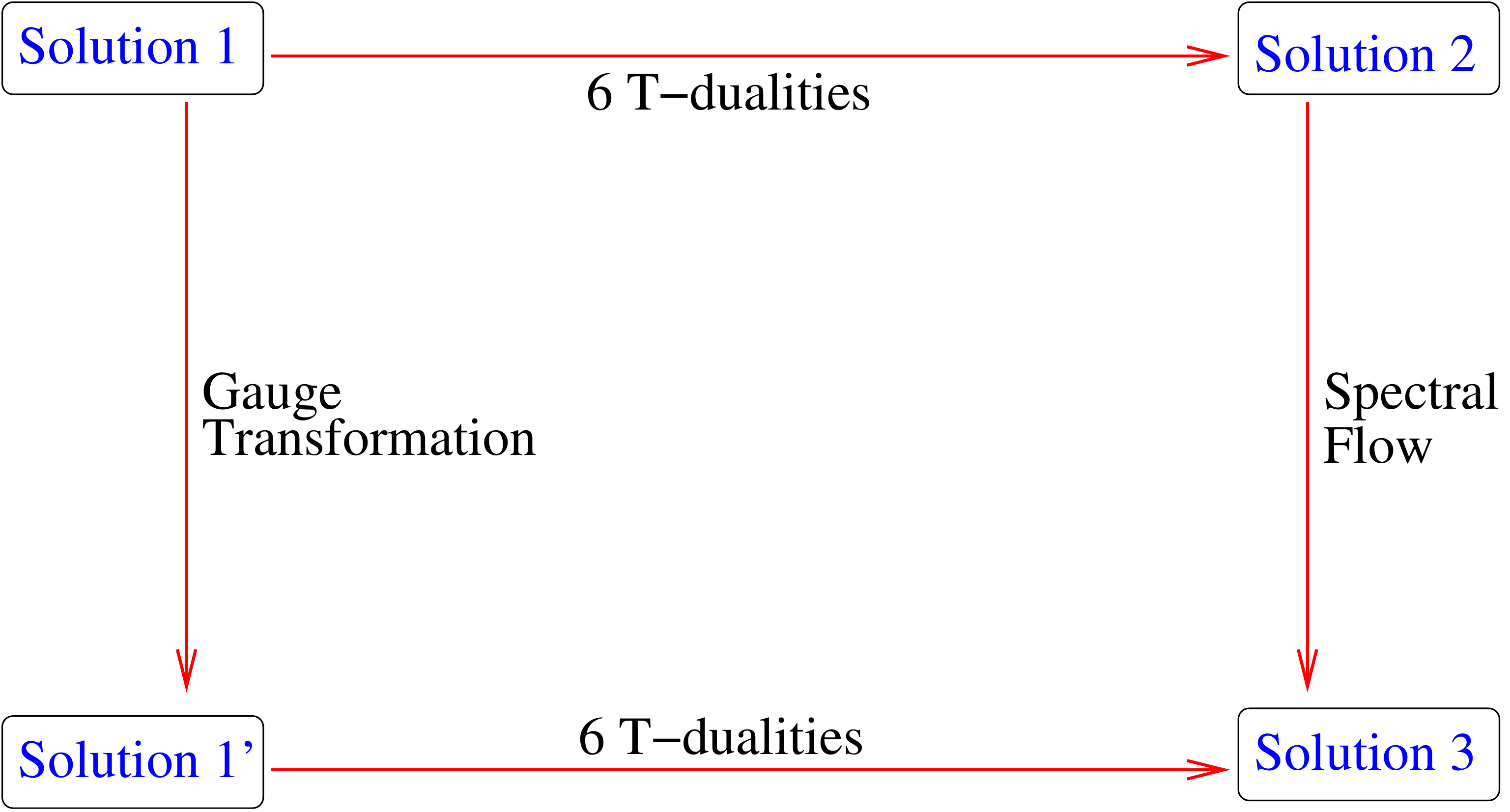}
\caption{\footnotesize We depict the commutative diagram expressing the link between spectral flow, large gauge transformations and T-duality. If one starts from a given BPS solution (solution 1), one can perform 6 T-dualities on it to obtain solution 2. On the other hand, one can first perform a large gauge transformation \eqref{ggetsfoBPS} on solution 1 to obtain solution 1', that only differs from solution 1 by the values of the B-field Wilson lines. If one then perform 6 T-dualities on solution 1', one obtains solution 3. Solution 2 and solution 3 are related by a spectral flow transformation \cite{Bena:2008wt}.}
\end{figure}

\section{T-dualities and the almost BPS solutions} \label{galalBPS}

As we have shown in eq. (\ref{bpst6}), performing 6 T-dualities on a BPS solution simply exchanges the various harmonic functions defining the solution. Hence, after such an operation we end up in the same class of solutions we started from. The situation in the almost BPS case is not as simple, but also much richer: the subset of almost BPS solutions is not closed under T-duality, and thus we can use T-duality to construct new classes of solutions.

We present in this section the general solution obtained by applying 6 T-dualities on a generic almost-BPS solution, and we will discuss several interesting particular cases in section \ref{subcases}. Since almost BPS solutions cannot be written in general in a closed form, we will not write explicit expressions for the $Z_I$ and $\mu$ functions, but use their implicit definitions given via the differential equations (\ref{nonBPSeq}). We will use the harmonic functions $V$ and $P_I = K_I$ to describe the 4D base metric and the dipole gauge fields $a_I$.

Buscher's  rules reviewed in section \ref{sec:Tdualities} give us the type IIA solution obtained by applying 6 T-dualities on the almost-BPS solution of section \ref{sec:thesetup}. After dualities, the NSNS fields are given by
\begin{eqnarray}
 ds_{10}^2 &=& -e^{2U} \, (dt + \omega)^2 + e^{-2U} \, ds_3^2 + \sum_{I=1}^3 \, \frac{e^{-2U}}{Z_I \Delta_I V} ds_{I}^2 \,, \nonumber \\ 
 \rm{e}^{-2\Phi} &=& e^{6U}{Z^3 \Delta^3 V^3} \,, \\[2mm]
B^{(2)} &=& -\sum_{I=1}^3\frac{1}{\Delta_I}\Bigl({K_I}-{\mu\over Z_I} \Bigr) dT_I \,, \nonumber 
\end{eqnarray}
where we recall that the $\Delta_I$ are given by \eqref{NSNStsfo}:
\begin{eqnarray}
\Delta_I = G_I^2+B_I^2 = {C_{IJK}\over 2} {Z_J Z_K\over V Z_I } + K_I^2 - 2 {K_I \mu \over Z_I }
\end{eqnarray}
and we have defined $\Delta = (\Delta_1 \Delta_2 \Delta_3)^{1/3}$. 
The RR-fields are  
\begin{eqnarray}
C^{(1)} \!\!&=&\!\! e^{4U}\left(Z^3 - V \mu Z_I K_I +V \sum_{I<J}K_I Z_I K_J Z_J -K^3 V^2 \mu \right)(dt+\omega) + v_0 \,, \nonumber \\[2mm] 
C^{(3)}_I \!\!&=&\!\!  \frac{1}{V Z_I \Delta_I}\left(\sum_J (Z_J K_J -2 Z_I K_I -K^3 V)(dt+\omega)\right) - v_I - \frac{1}{\Delta_I}\left(K_I -\frac{\mu}{Z_I}\right)v_0  \,, \nonumber \\[3mm] 
C^{(5)}_{JK} \!\!&=&\!\! \frac{1}{V^2 Z_J Z_K \Delta_J \Delta_K} \Big[ (K_J K_K V + C_{IJK}Z_I) (Z_J Z_K + C_{IJK}K_I V \mu)  \\[4mm] \nonumber 
&& \quad \quad - V ( K_J Z_J + K_K Z_K )(C_{IJK}K_I Z_I + \mu)  \Big] (dt+ \omega)  -C_{IJK}w^I  \\[3mm] \nonumber 
&&\!\! + \frac{1}{\Delta_J}\!  \left(K_J - \frac{\mu}{Z_J}\right)\! v_K + \frac{1}{\Delta_K}\!  \left(K_K - \frac{\mu}{Z_K}\right)\! v_J  +\frac{1}{\Delta_J \Delta_K}\! \left(K_J - \frac{\mu}{Z_J}\right) \! \left(K_K - \frac{\mu}{Z_K}\right) v_0 \,, \\[4mm] \nonumber
C^{(7)} \!\!&=&\!\!\Big[ \frac{1}{V^2 Z^3 \Delta^3} \big( 2Z^3 \mu - (K_I Z_I + K^3 V)(Z^3 + V \mu^2) +2 V\mu \sum_{I<J}Z_I K_I Z_J K_J \big)(dt + \omega) \\[3mm] \nonumber 
&& + A + \frac{1}{\Delta_I}\left(K_I - \frac{\mu}{Z_I}\right) w^I - \frac{C_{IJK}}{2} \frac{1}{\Delta_J \Delta_K}\left(K_J - \frac{\mu}{Z_J}\right)\left(K_K - \frac{\mu}{Z_K}\right) v_I  \\[4mm] \nonumber 
&& -\frac{1}{\Delta^3} \frac{C_{IJK}}{6}\left(K_I - \frac{\mu}{Z_I}\right)\left(K_J - \frac{\mu}{Z_J}\right)\left(K_K - \frac{\mu}{Z_K}\right) v_0 \Big] \wedge dT_1 \wedge dT_2 \wedge dT_3 \,, 
\end{eqnarray}
where we used once more that $K=(K_1K_2K_3)^{1/3}$. 
In the language of 11-dimensional supergravity, the above solution can be recast in the simpler form
\begin{eqnarray}
\label{new11Dfields}
ds_{11}^2 &=& \frac{e^{-4U}}{Z^2 \Delta^2 V^2} \left[ (d\psi + v_0 ) + C^{(1)}_t (dt + \omega) \right]^2 -e^{4U} Z \Delta V (dt + \omega)^2 + Z \Delta V ds_3^2 \nonumber \\[2mm] 
&& + \sum_I \frac{Z \Delta}{Z_I \Delta_I} ds_I^2 \,, \\[2mm]
A^{(3)} &=& \sum_I \left[ \frac{1}{V Z_I \Delta_I}\left(\sum_{J \neq I} Z_J K_J - Z_I K_I -K^3 V\right)(dt+\omega) \right. \nonumber \\[4mm]
&& \left.-\frac{1}{\Delta_I}\left( K_I - \frac{\mu}{Z_I} \right) (d\psi + v_0) - v_I \right] \wedge dT_I \,.
\end{eqnarray}

This general class of solutions can be split in two broad subclasses: those solutions for which 
 $g_{tt}$ vanishes and those for which it does not.  This is controlled by the quantity
\begin{eqnarray}
\label{defF}
 F^2 &=& e^{4U} Z^3 \Delta^3 V^3 - e^{-4U} {C^{(1)}_t}^2  \\[2mm] \nonumber
 &=& K_I^2 Z_I^2 -2 \sum_{I<J} K_I Z_I K_J Z_J - 2( K_I Z_I ) K^3 V + K^6 V^2 + 8 K^3 V \mu \,,
\end{eqnarray}
in terms of which one can write
\begin{eqnarray}
g_{tt}=-F^2/(Z \Delta V)^2\,.
 \end{eqnarray}
Of course the metric component $g_{tt}$ vanishes when $F$ does. 
We will consider the case of vanishing $g_{tt}$ in the following, so let us first concentrate on the generic situation in which $g_{tt}$ is non-zero. 
In this case, $F^2$ has to be positive for $\partial_t$ to be timelike. 
Completing the squares with respect to $dt$
\begin{eqnarray}
 ds_{11}^2 &=& -\frac{F^2}{Z^2 \Delta^2 V^2} \left(dt + \omega - \frac{e^{-4U} C^{(1)}_t}{F^2}( d\psi +v_0) \right)^2 + \frac{Z \Delta V}{F^2} (d\psi + v_0 )^2  \nonumber \\[2mm]
 && + Z \Delta Vds_3^2 + \sum_I \frac{Z \Delta}{Z_I \Delta_I} ds_I^2 \,,
\end{eqnarray}
we can rewrite the 11-dimensional metric and 3-form gauge field in a form identical to \eqref{11Dfields}
\begin{eqnarray}
\label{new11Dmet}
ds_{11}^2 &=& -\widetilde{Z}^{-2}\left(dt+\widetilde{k}\right)^2+\widetilde{Z} d{\tilde s}^2_4+\sum_{I=1}^3{\widetilde{Z}\over \widetilde{Z}_I} ds^2_I\,,\nonumber\\
A^{(3)}  &=& \sum_I \left[ -\frac{1}{\widetilde{W}_I} (dt+\widetilde{k}) + \widetilde{a}_I \right] \wedge dT_I \,.
\end{eqnarray}
The new 4-dimensional base is
\begin{eqnarray}
\label{new4Dmet}
d{\tilde s}^2_4 &=& \widetilde{V}^{-1}(d\psi+\widetilde{A})^2+ \widetilde{V} ds^2_3 \quad {\rm{with}} \quad \widetilde{V} = F \quad {\rm and} \quad \widetilde{A} = v_0 \,,
\end{eqnarray}
where the three-dimensional metric $ds_3^2$ is flat. The various  other quantities describing the solution (\ref{new11Dmet}) are given by
\begin{eqnarray}
\label{new11Ddatas}
 \widetilde{Z}_I &=& \frac{Z_I \Delta_I V}{F} \,,\quad \widetilde{Z}=(\widetilde{Z}_1 \widetilde{Z}_2 \widetilde{Z}_3)^{1/3} \\
 \widetilde{k} &=& \widetilde{\mu}(d\psi+\widetilde{A}) + \tilde{\omega}\,,\\
 \widetilde{\mu} &=&\frac{1}{F^2} \,\left(-Z^3 + V \mu Z_I K_I - V \sum_{I<J}K_I Z_I K_J Z_J + K^3 V^2 \mu\right) \,, \quad \tilde \omega = \omega\,,\\
 \widetilde{W}_I &=& -\frac{V Z_I \Delta_I}{\sum_{J \neq I} Z_J K_J - Z_I K_I - K^3 V}\,, \\
 \widetilde{a}_I &=&  \widetilde{P}_I (d\psi + \widetilde{A}) + \widetilde{w}^I \,,\\
 \widetilde{P}_I &=& -\frac{1}{\Delta_I} \left( K_I - \frac{\mu}{Z_I} \right) + \frac{\widetilde{\mu}}{\widetilde{W}_I} \,, \quad \widetilde{w}^I = -v_I \,.
\end{eqnarray}

From the explicit expressions given above, we can make two general observations.
First, the functions $\widetilde{W}_I$ that appear in the time component of the gauge field $A^{(3)}$, and thus encode the M2-brane charges of the solution, are in general not equal to the functions
$\widetilde{Z}_I$ that give the warp factors of the metric $ds^2_{11}$. This is unlike our original 
11-dimensional ansatz (\ref{11Dfields}), that was characterized by the fact that the {\it same} functions $Z_I$ appeared both in the metric and in the gauge field. Solutions of this latter kind have the property that a probe M2-brane feels no force when placed in such a geometry: as explained in \cite{Bena:2009fi}, where an ansatz of the type (\ref{11Dfields}) was denoted by the name ``floating brane'' ansatz, this property implies crucial simplifications in the construction of the solutions. As we have just seen, generically  T-duality maps ``floating brane'' solutions into ``non-floating brane'' ones, and could thus provide important clues for the construction of a more general class of non-BPS geometries.  

Secondly, the 4D base metric one obtains after 6 T-dualities (\ref{new4Dmet}), like the original Gibbons-Hawking metric (\ref{GHmetric}), has the form of a U(1)-fibration over a {\it flat} 3-dimensional space: however the quantities $\widetilde V=F$ and $\tilde A= v_0$ that describe this fibration are far from being of the Gibbons-Hawking type, generically. This is another manifestation of the fact that T-dualities produce solutions that lie in a much larger class than the original almost-BPS class. We will leave the task of analyzing further the generic solutions in (\ref{new11Dmet}) for the future, and in the following we will concentrate on some simpler particular cases.

\section{T-dualities, gauge transformations and spectral flows in the non-BPS case}  \label{SFggetsfononBPS}

We have seen in section \ref{gaugeSFBPS} that, in the supersymmetric case, spectral flow transformations and large gauge transformations are related by T-dualities, a relation that is summarized in eq. (\ref{spectralgaugelink}).

We investigate in this section whether the relation found in the supersymmetric case holds in the non-supersymmetric one as well. 

We face however a technical difficulty: The action of one spectral flow transformation on almost-BPS solutions was derived in \cite{Bena:2009fi}. In principle one could iterate the calculation of
\cite{Bena:2009fi} to find how almost-BPS solutions transform under a general (triple) spectral flow. This computation however seems too cumbersome to be carried out explicitly. We thus do not know how to compute the left hand side of eq. (\ref{spectralgaugelink}) on almost-BPS solutions. The best we can do, for almost-BPS solutions, is to compute the right hand side of  (\ref{spectralgaugelink}) and perform some partial checks on the identity. Turning things around, we can assume the validity of (\ref{spectralgaugelink}) when applied to almost-BPS solutions and use it to derive the action of a general spectral flow transformation on almost-BPS geometries.

 Let us then compute the action of the left hand side of (\ref{spectralgaugelink}) on an almost-BPS solution. In section \ref{galalBPS}, we have obtained the general form of the geometry resulting from the action of 6 T-dualities on an almost BPS solution. For our purposes we need the inverse of this transformation, which, as can be easily checked, is given by itself, up to reversing the sign of all the RR fields. We can then perform on this solution a large gauge transformation, which amounts to shift $B^{(2)}$ by the constant term $-\gamma_I d T_I$, giving
\begin{eqnarray}
B^{(2)} =  -\sum_{I=1}^3 \left[ \frac{1}{\Delta_I}\Bigl({K_I}-{\mu\over Z_I} \Bigr) + \gamma_I \right] dT_I \,, 
\end{eqnarray}
where we recall that the $\Delta_I$ are given by \eqref{NSNStsfo}:
\begin{eqnarray}
\Delta_I = G_I^2+B_I^2 =  {C_{IJK}\over 2} {Z_J Z_K\over V Z_I } + K_I^2 - 2 {K_I \mu \over Z_I }\,.
\end{eqnarray}
We can now apply one more time the rules for 6 T-dualities. The final solution is 
\begin{eqnarray}
 ds_{10}^2 &=& - e^{2U} ( dt + \omega )^2 + e^{-2U} ds_3^2 + \sum_I \frac{e^{-2U}}{N_I} ds_I^2 \,, \nonumber \\
 e^{-2\phi} &=& e^{6U} N^3\,,  \\[2mm]
 \quad B^{(2)} &=& -\frac{V Z_I}{N_I} \left( K_I - \frac{\mu}{Z_I} +\frac{1}{\gamma_I} \right) +\frac{1}{\gamma_I} \,, \nonumber
\end{eqnarray}
\begin{eqnarray}
 C^{(1)} \!\!&=&\!\! e^{4U}\Big[ -T^3 \, V^2 \mu + \frac{C_{IJK}}{2} \left( \gamma_I \, V T_J Z_J T_K Z_K \right) - \frac{C_{IJK}}{2} \gamma_J \gamma_K \, T_I Z_I V \mu + \gamma^3 \, Z^3 \Big] ( dt+ \omega ) \nonumber \\
 &&\!\! + A - \gamma_I \, w^I - \frac{C_{IJK}}{2} \gamma_J \gamma_K \, v_I + \gamma^3 \, v_0 \,, \\ \nonumber
 C^{(3)}_I \!\!&=&\!\! \frac{1}{N_I} \left( -T^3 V - \frac{C_{IJK}}{2} \gamma_J \gamma_K \, T_I Z_I + C_{IJK} \gamma_I \gamma_J \, T_K Z_K  \right) (dt + \omega ) \\ \nonumber
 && \!\! \hspace{-1.4cm}+ \frac{1}{N_I} \left[ T_I Z_I V\! \left(K_I - \frac{\mu}{Z_I}\right) \! + \gamma_I \left(\frac{C_{IJK}}{2}Z_J Z_K - K_I V \mu \right)\! \right]\! \! \left(\! A - \gamma_I \, w^I - \frac{C_{IJK}}{2} \gamma_J \gamma_K \, v_I + \gamma^3 \, v_0\! \right) \\ \nonumber 
 && + w^I + C_{IJK} \gamma_J  \, v_K - \frac{C_{IJK}}{2} \gamma_J \gamma_K \, v_0 \,.
\end{eqnarray}
We have defined the new functions
\begin{eqnarray}
 T_I = 1 + \gamma_I \, K_I \,,\quad N_I = \frac{C_{IJK}}{2} \gamma_I^2 \, Z_J Z_K + V T_I^2 Z_I - 2 \gamma_I \, V T_I \mu  \,,
\end{eqnarray}
and, as usual, we use the short-hand notations $K^3=K_1 K_2 K_3$, $Z^3= Z_1 Z_2 Z_3$, $\gamma^3=\gamma_1 \gamma_2 \gamma_3$, $T^3 = T_1 T_2 T_3$ and $N^3=N_1 N_2 N_3$.

The 11-dimensional lift of this solution writes as
\begin{eqnarray}
\label{galsolution}
ds_{11}^2 &=& -\widetilde{Z}^{-2} \, \left( dt + \widetilde{k} \right)^2 + \widetilde{Z} \, ds^2_4 + \sum_{I}{\widetilde{Z} \over \widetilde{Z}_I} \, ds^2_I \,, \\
A^{(3)} &=& \sum_I \left[ -{ 1 \over \widetilde{W}_I } ( dt + \widetilde{k} ) + \widetilde{P}_I \, (d\psi + \widetilde{A})+ \widetilde{w}^I \right ] \wedge dT_I \,,
\end{eqnarray}
where
\begin{eqnarray}
\label{4Dnewmet}
 ds_4^2 \!\!&=&\!\! \widetilde{V}^{-1}(d\psi + \widetilde{A})^2 + \widetilde{V} ds_3^2 \quad \rm{with} \nonumber \\[2mm]
 \widetilde{V} \!\!&=&\!\! \left[ {C_{IJK} \over 2} \gamma_J^2 \gamma_K^2 T_I^2 Z_I^2 - C_{IJK} \gamma_I^2 \gamma_J \gamma_K T_J Z_J T_K Z_K\right. \\[2mm] \nonumber
&& \quad \quad \quad\left. - T^3 V (C_{IJK} \gamma_J \gamma_K T_I Z_I) + T^6 V^2 + 8 \gamma^3 T^3 V \mu \right]^{1/2}  \,, \\[2mm] \nonumber 
 \widetilde{A} \!\!&=&\!\! A - \gamma_I \, w^I - \frac{C_{IJK}}{2} \gamma_J \gamma_K \, v_I + \gamma^3 \, v_0
\end{eqnarray}
and
\begin{eqnarray}
\label{newdatas}
 \widetilde{Z}_I &=& \frac{N_I}{\widetilde{V}} \,, \\
 \widetilde{k} &=& \widetilde{\mu}(d\psi+\widetilde{A}) + \omega \quad \rm{with} \\[2mm]
 \widetilde{\mu} &=&  \frac{1}{\widetilde{V}^2}\left( -\gamma^3 \, Z^3 + {C_{IJK} \over 2} \gamma_J \gamma_K \, Z_I T_I V \mu - {C_{IJK} \over 2} \gamma_I \, V T_J Z_J T_K Z_K + T^3 V^2 \mu \right) , \\[3mm]
 \widetilde{W}_I &=& \frac{N_I}{T^3 V + \frac{C_{IJK}}{2}\gamma_J \gamma_K \, T_I Z_I - C_{IJK} \gamma_I \gamma_J \, T_K Z_K} \,, \\[3mm]
 \widetilde{P}_I &=&  \frac{1}{N_I} \Big( V Z_I T_I K_I + \frac{C_{IJK}}{2} \gamma_I \, Z_J Z_K -(2 T_1 -1 ) V \mu \Big) + \frac{\widetilde{\mu}}{\widetilde{W}_I}\,, \\[2mm]
 \widetilde{w}^I &=& w^I + C_{IJK} \gamma_J \, v_K - \frac{C_{IJK}}{2} \gamma_J \gamma_K \, v_0 \,
\end{eqnarray}
and $\widetilde{Z}=(\widetilde{Z}_1 \widetilde{Z}_2 \widetilde{Z}_3)^{1/3}$. 

We can now compare the solution above with the one produced by a spectral flow transformation on an almost-BPS solution, and thus verify the relation (\ref{spectralgaugelink}). As we explained before, we can only do this explicitly when a single spectral flow parameter is non-vanishing.
Thanks to the symmetry of  the solution in \eqref{4Dnewmet}-\eqref{newdatas} with respect to the indices $I=1,2,3$ we can pick the non-vanishing parameter to be $\gamma_1$. Setting $\gamma_2=\gamma_3=0$ in  \eqref{newdatas} we find
\begin{eqnarray}
 \widetilde{V} &=& T_3 \, V  \,,  \quad \widetilde{A} = A - \gamma_3 \, w^3 \,, \nonumber \\[2mm] \nonumber 
 \widetilde{Z}_1 &=& \widetilde{W}_1 = \frac{Z_1}{T_3} \,, \quad \widetilde{Z}_2 = \widetilde{W}_2= \frac{Z_2}{T_3} \,, \\[2mm]
 \widetilde{Z}_3 &=& \widetilde{W}_3 = \frac{P_3}{V T_3} \,, \quad \widetilde{\mu} =  \frac{\mu}{T_3} - \gamma_3 \, \frac{Z_1 Z_2}{V T_3^2} \,, \\[2mm] \nonumber 
 \widetilde{P}_1 &=& K_1 - \gamma_3 \, \frac{Z_2}{V T_3} \,, \quad  \widetilde{P}_2 = K_2 - \gamma_3 \, \frac{Z_1}{V T_3} \,, \quad  \widetilde{P}_3 = \frac{K_3}{T_3} \,, \\[2mm]
 \widetilde{w}^1 &=& w^1 +\gamma_3 \, v_2 \,, \quad \widetilde{w}^2 = w^2 +\gamma_3 \, v_1 \,, \quad \widetilde{w}^3 = w^3 \,. \nonumber
\end{eqnarray}
These are exactly the data describing the solution obtained in \cite{Bena:2009fi} from one spectral flow. 
Although this is an important non-trivial check of the relation (\ref{spectralgaugelink}) in the almost-BPS case, it, of course, leaves open the possibility that the relation might be violated by terms of order higher than linear in the $\gamma_I$'s. We think this possibility, however, to be quite unlikely. As we have already recalled, spectral flow is a well defined element of the U-duality group, $G=$SU(1,1)$^3$, of the $STU$ model, resulting from the composition of a certain U-duality transformation and a rotation inside $T^6\times S^1_\psi$. The relation (\ref{spectralgaugelink}) expresses a quite non-obvious identity between two, a priori different, elements of $G$. It is possible that this relation, as an identity in $G$, be actually corrected as 
\begin{equation}
\mathcal{F}(\gamma_I)=\mathcal{T}_6 \circ \mathcal{G}(\gamma_I) \circ \mathcal{T}_6^{-1}\circ \mathcal{R}(\gamma_I)\,,
\end{equation}
with $\mathcal{R}(\gamma_I)$ some element in $G$, symmetric in the parameters $\gamma_I$'s. We have shown, however, that $\mathcal{R}(\gamma_I)=\mathrm{id}$ when $\gamma_2=\gamma_3=0$ and when acting on BPS solutions: hence we advance the natural conjecture that
$\mathcal{R}(\gamma_I)=\mathrm{id}$ {\it identically} in $G$.

It might be useful to note that  the solution \eqref{galsolution}-\eqref{newdatas} contains, as particular limits, both the original almost-BPS solution \eqref{11Dfields}-\eqref{nonBPSeq} and the solution \eqref{new11Dmet}-\eqref{new11Ddatas} obtained from it after 6 T-dualities. The general solution \eqref{galsolution}-\eqref{newdatas} obviously reduces to the almost-BPS solution when 
the gauge transformation parameters are set to zero:  $\gamma_I=0$, for $I=1,2,3$. It is a bit more subtle to recover the T-dualized solution \eqref{new11Dmet}-\eqref{new11Ddatas}: for this purpose one has to take the limit of large $\gamma_I$'s, i.e. $\gamma_1=\gamma_2=\gamma_3=\gamma\to \infty$. Let us make this point more precise: Remember that the solution  \eqref{galsolution}-\eqref{newdatas} has been obtained by first performing 6 T-dualities, followed by a large gauge transformation and again 6 T-dualities. We look for a limit in which the last two steps essentially reduce to the identity map. After a large gauge transformation 
\begin{eqnarray}
 B_I \to B_I - \gamma
\end{eqnarray}
the factors $\Delta_I$ that appear in T-duality transformation rules become
\begin{eqnarray}
 \Delta_I = \gamma^2 - 2 \gamma B_I + (B_I^2 + G_I^2)\,.
\end{eqnarray}
From the T-duality map given in section \ref{sec:Tdualities}, one then sees that under the combination of a gauge transformation and 6 T-dualities, and in the large $\gamma$ limit, 
the NSNS fields transform as
\begin{eqnarray}
 {\rm{e}}^{2\widetilde{\phi}} \sim \frac{{\rm{e}}^{2\phi}}{\gamma^6} \,, \quad \widetilde{G}_I \sim \frac{G_I}{\gamma^2} \,, \quad \widetilde{B}_I \sim \frac{1}{\gamma} + \frac{B_I}{\gamma^2}\,.
\end{eqnarray}
Thus, up to a renormalization by the appropriate factors of $\gamma$, this transformation reduces to the identity map on the dilaton and the metric components $G_I$'s.  The B-field components $B_I$'s, because this is where the gauge transformations acted in the first step, behave a bit differently: 
at leading order in $1/\gamma$ they just reduce to a constant term $1/\gamma$, and 
one has to keep the next to leading corrections, of order $1/\gamma^2$, to recover the original $B_I$'s. From the transformation rules for the RR-fields \eqref{6Ttsfo}, one sees that for $\gamma\to \infty$:
\begin{eqnarray}
 \widetilde{C}^{(1)} \sim \gamma^3 C^{(1)} \,, \quad \widetilde{C}^{(3)}_I \sim \gamma \, C^{(3)}_I \,, \quad \widetilde{C}^{(5)}_{JK} \sim \gamma^{-1} C^{(5)}_{JK} \,, \quad \widetilde{C}^{(7)} \sim \gamma^{-3} C^{(7)} \,,
\end{eqnarray}
and thus even on RR fields the map under consideration reduces to the identity, up  renormalization by appropriate factors of $\gamma$.  One concludes that in the large $\gamma$ limit the series of transformations that lead to the solution \eqref{galsolution}-\eqref{newdatas} reduces to just the 6 T-duality transformation of section \ref{sec:Tdualities}, and the solution
\eqref{galsolution}-\eqref{newdatas} reduces to  \eqref{new11Dmet}-\eqref{new11Ddatas}.

To summarize, the geometry given in eqs. \eqref{galsolution}-\eqref{newdatas} is the most general solution we have obtained so far. We will examine several interesting subcases in section 
\ref{subcases}.

\subsection{Generic Calabi--Yau compactifications}

As we already mentioned previously, the internal space we used to reduce the theory to 5 dimensions can be considered as a simple prototype of a generic Calabi--Yau manifold.
In fact, $T^6/Z_2 \times Z_2$ is a Ricci-flat SU(3) holonomy space (actually, the holonomy is $Z_2 \times Z_2 \subset $ SU(3)).
The 5- and 4-dimensional supergravity models resulting from such a compactification process are then specified by the cubic scalar couplings defined by the $C_{IJK}$ coefficients.
In our example (the STU model), these coefficients are simply $C_{IJK} =|\epsilon_{IJK}|$, where $I = 1,2,3$.
However, one could rather easily replace our internal space with a generic Calabi--Yau manifold and still solve all the equations of motion.
In the generic case, the $C_{IJK}$ coefficient are going to be related to the triple intersection numbers of the manifold and the index $I = 1,...,h_{(1,1)}$ labels the number of 2-cycles on the manifold, which, for M-theory/IIA reductions, give the number of vector multiplets $n_V$.

For a general Calabi--Yau manifold, the resulting 4-dimensional model will not have all the symmetries that are present in the special STU case.
However, it was shown \cite{Ferrara:1988ff,Bodner:1990zm,deWit:1992wf} that any such compactification will have at least $n_V + 1$ residual symmetries, related to the isometries of the vector multiplet scalar manifold.
One of these is the 4-dimensional S-duality transformation, while the remaining $n_V$ ones can be identified with the Peccei--Quinn symmetries of the axion fields in each vector multiplet.
These transformations are precisely the ones we used in our work, because, as we will see more in detail later, the combined action of the 6 T-duality transformations can be identified with the 4-dimensional S-duality action exchanging each scalar field with its inverse and the gauge transformations described by the $\gamma_I$ parameters are identified with the shift symmetries of the axions.
We therefore conclude that the final solution we constructed should give the most general non-BPS multicentre configuration for a generic Calabi--Yau compactification.

\section{Interesting subcases after 6 T-dualities} \label{subcases}

In this section, we examine particular cases of the non-BPS solutions 
\eqref{new11Dmet}-\eqref{new11Ddatas},  that are simple enough to be analyzed in detail.

\subsection{From black holes to black strings} \label{KKK=0}

\subsubsection{A non BPS black string solution}

The simplest almost-BPS solutions are the ones without any magnetic fields: $K_1=K_2=K_3=0$, $w^1=w^2=w^3=0$. They describe single center non-BPS rotating black holes (or multiple non-interacting rotating black holes). In the absence of magnetic fields, the solution drastically simplifies and can be written in a closed form in terms of five harmonic functions $V$, $M$ and $L_I$, $I=1,2,3$. The single center black hole of this kind was analyzed in detail in \cite{Bena:2009ev}: It is given by a geometry of the form \eqref{11Dfields}, where the various quantities have the following explicit expressions (written in polar coordinates on $\mathbb{R}^3$):
\begin{eqnarray}
 V &=& 1 + \frac{Q_6}{r} \,, \quad A = -Q_6 \cos\theta d\phi \,, \nonumber \\ 
 K_1 &=& K_2 = K_3 = 0 \,, \quad Z_I = L_I = 1 + \frac{Q_I}{r} \,, \\ \nonumber 
 M &=&  m_0 + \frac{m}{r} + \alpha \frac{\cos\theta}{r^2} \,, \quad \mu = \frac{M}{V} \,, \quad \omega = m \cos\theta d\phi - \alpha \frac{\sin^2\theta}{r} d\phi \,.
\end{eqnarray}
Regularity requires $m$ to be zero. This ensures that the metric be regular for the range of parameters for which the warp factor
\begin{eqnarray}
 e^{-4U} = L_1 L_2 L_3 V - M^2 \,,
\end{eqnarray}
is everywhere positive (i.e. for $1-m_0^2>0$ and $Q_1 Q_2 Q_3 Q_6 -\alpha^2>0$).

We want to consider the geometry obtained by applying 6 T-dualities on the above solution. The analogous operation in the BPS case produced the BPS black string of section \ref{bpsstring:sec}. We thus expect to find by this method a non-BPS black string. 

The 10-dimensional solution, that can be easily obtained as a particular case of the solution of section \ref{galalBPS} is given by
\begin{eqnarray}
\label{10DBSTR}
 ds_{10}^2 &=& -e^{2U} \, (dt + \omega)^2 + e^{-2U} \, ds_3^2 + \frac{e^{-2U}}{L^3} \sum_{I=1}^3 \,  L_I ds_{I}^2 \,, \nonumber \\ 
 \rm{e}^{-2\Phi} &=& (e^U L)^6 \,, \\[2mm]
B^{(2)} &=& \frac{M}{L^3} \sum_{I=1}^3 L_I dT_I \,, \nonumber 
\end{eqnarray}
and 
\begin{eqnarray}
C^{(1)} &=& { L^3 e^{4U} } (dt+\omega) \,, \nonumber \\
C^{(3)} &=& -\sum_{I=1}^3 v_I \wedge dT_I \,, \nonumber \\
C^{(5)}_{JK} &=& C_{IJK}\frac{1}{L_I} (dt+\omega) - \frac{M}{L^3} \left( L_J v_K + L_K v_J \right) \,,  \\[2mm]
C^{(7)} &=& \left( {2M \over L^3} (dt+\omega) -\frac{M^2}{L^3}\sum_I\frac{v_I}{L_I} + A \right) \wedge dT_1 \wedge dT_2 \wedge dT_3\,, \nonumber
\end{eqnarray}
where we defined $L = (L_1 L_2 L_3)^{1/3}$, and we recall that $\star d v_I = dL_I$.

The 11-dimensional lift of this solution produces a geometry for which $g_{tt}=0$ and, hence, which does not fall in the generic class of solutions described in section \ref{galalBPS}. The 11-dimensional configuration one obtains is instead given by
\begin{eqnarray}
\label{Bstrsol}
 ds_{11}^2 &=& \frac{2}{L}(dt+\omega)d\psi + \frac{e^{-4U}}{L^4}d\psi^2 + L^2 ds_3^2 + \sum_I\left(\frac{L_I}{L} ds_I^2\right) \,, \\[2mm]
 A^{(3)} &=& \sum_I \left( \frac{M}{L^3} L_I d\psi -v_I \right) \wedge dT_I \,.
\end{eqnarray}
In the following we discuss the physical properties of this solution, and show that it indeed describes a non-BPS black string. 

\subsubsection{Physical properties}

Let us first investigate the regularity of the solution. Absence of Dirac-Misner strings requires the $\phi$ component of 1-form $\omega$ to vanish on the $\phi$-axis of rotation, i.e. at
both at $\theta=0$ and $\theta=\pi$; hence, as in the solution before T-dualities, one must set 
\begin{eqnarray}
 m = 0 \,.
\end{eqnarray}
One can moreover verify that in the allowed range of parameters ($Q_1 Q_2 Q_3 Q_6 -\alpha^2>0$ and $1-m_0^2>0$) $e^{-4 U}$ and $e^{-4U}r^2\sin^2\theta -\omega^2$ are always positive, as is required for the absence of closed time-like curves.

Let us now look at the asymptotic (large $r$) limit of the metric:
\begin{eqnarray}
 ds_{\infty}^2 =  2 dt d\psi + (1-m_0^2) d\psi^2 + ds_3^2 + \sum_I ds_I^2\,.
\end{eqnarray}
In this form the metric does not appear to be asymptotically flat, but this can be easily remedied by the change of coordinates
\begin{eqnarray}
 d\psi \to \sqrt{1-m_0^2} \left( d \psi -\frac{1}{1-m_0^2} dt \right) \,, \quad dt \to \frac{1}{\sqrt{1-m_0^2}} dt \,,
\end{eqnarray}
which brings the metric to the form
\begin{eqnarray}
 ds_{\infty}^2 =  -dt^2 + d\psi^2 + ds_3^2 + \sum_I ds_I^2\,.
\end{eqnarray}
Assuming the identifications $(\phi,\psi) \sim (\phi + 2\pi, \psi)\sim (\phi,\psi + 2\pi R_\psi)$, for an arbitrary $R_{\psi}$, this is the regular flat metric of ${\mathbb R}^{1,3}\times S^1 \times T^6$. 

One can now compute the charges of the solution. We define the asymptotic charges as the integrals of the RR field strengths $F^{(i+1)}$ given in \eqref{fieldstrength}-\eqref{fieldstrengthbis} over cycles of the form $S^2_\infty\times T^{i-1}$ ($i=1,3,5,7$), with $S^2_\infty$ the two sphere at asymptotic infinity in $\mathbb{R}^3$. Note however that the Chern-Simons terms $H^{(3)}\wedge C^{(i-2)}$ do not contribute to the integrals. Hence the charges are given by
\begin{eqnarray}
\label{chgesBSTR}
 Q^{D6} &=& \int_{S^2_{\infty}} d C^{(1)} = 0 \,, \nonumber \\[3mm]
 Q^{D4}_{JK} &=& \int_{S^2_{\infty}\times T_I^2} d C^{(3)} = C_{IJK} Q_I \,, \\[3mm]
 Q^{D2}_I &=& \int_{S^2_{\infty}\times T_{JK}^4} d C^{(5)} = m_0 \,C_{IJK} Q_J  \,, \nonumber \\[3mm] \nonumber
 Q^{D0} &=& \int_{S^2_{\infty}\times T^6} d C^{(7)} = Q_6 + m_0^2 ( Q_1 + Q_2 + Q_3 ) \,.
\end{eqnarray}
The solution also has angular momentum in the ${\mathbb R}^3$ space transverse to the string: 
it is encoded in the large $r$ limit of the 1-form $\omega$, giving the $dt d\phi$ term of the 10D metric, and it is equal to
\begin{equation}
 J=\alpha \,.
\end{equation}
Let us finally examine the horizon geometry. The horizon is identified with the $t=\mathrm{const.}$, 
$z_{I,i}=\mathrm{const.}$, $r=0$ submanifold of the 11D metric. The metric induced on the horizon is:
\begin{eqnarray}
\label{hormetricBSTR}
 ds_{\rm{hor}}^2 &=& Q^2 d \theta^2 + Q^2\frac{Q_6 Q^3 - \alpha^2}{Q_6 Q^3 - \alpha^2\cos^2\theta} \sin^2 \theta d \phi^2 \\[2mm] \nonumber 
&& + \frac{Q_6 Q^3 - \alpha^2\cos^2\theta}{Q^4}\left(d\psi -\frac{Q^3 \alpha}{Q_6 Q^3 - \alpha^2\cos^2\theta} \sin^2\theta d\phi\right)^2
\end{eqnarray}
where we defined $Q=(Q_1Q_2Q_3)^{1/3}$. The space spanned by the coordinates $\theta,\phi$ is topologically, though not metrically, a 2-sphere (note that the coefficient of $d\phi^2$ becomes 
$Q^2 \sin^2\theta$ for $\theta=0,\pi$, a property which ensures the regularity of the horizon at the two poles of $S^2$). The coordinate $\psi$ identifies an $S^1$ whose radius varies over the $S^2$, but never vanishes. Though this $S^1$ is non-trivially fibered over the $S^2$, the fibration 1-form, $\frac{Q^3 \alpha}{Q_6 Q^3 - \alpha^2\cos^2\theta} \sin^2\theta d\phi$, is globally defined on $S^2$, and defines a topologically trivial U(1) bundle (this is a manifestation of the fact that the solution has vanishing D6 charge). One can thus conclude that the horizon has the topology of $S^2\times S^1$. As $\psi$ is a compact coordinate at infinity, this is really a black string.

It is easy to compute the area of the horizon manifold \eqref{hormetricBSTR},  from which one derives the entropy of the solution:
\begin{eqnarray}
S \propto \sqrt{Q_1 Q_2 Q_3 Q_6 - \alpha^2}\,.
\end{eqnarray}
This is a formula identical to the one giving the entropy of a black hole with charges $Q_1, Q_2, Q_3, Q_6$ and angular momentum $\alpha$: it might at first appear surprising that it also applies to our black ring solution, whose charges are given by \eqref{chgesBSTR} and are {\it not} to be confused with $Q_1, Q_2, Q_3, Q_6$. There is actually no contradiction. Indeed one can verify that the quartic invariant defined in \eqref{I4def}, evaluated for the charges
\begin{eqnarray}
p^0 =- \frac{1}{\sqrt{2}}Q^{D6}\,,\quad p^I = -\frac{C_{IJK}}{2\sqrt{2}} Q^{D4}_{JK}\,,\quad q_I =\frac{1}{\sqrt{2}} Q^{D2}_I\,,\quad q_0 =\sqrt{2} Q^{D0}\,,
\end{eqnarray}
reduces to 
\begin{eqnarray}
I_4(Q^{D6},Q^{D4},Q^{D2},Q^{D0}) = -\frac{1}{4}Q_1Q_2Q_3Q_6\,.
\end{eqnarray}
Therefore, the entropy is given by the expected formula
\begin{eqnarray}
S \propto \sqrt{-I_4(Q^{D6},Q^{D4},Q^{D2},Q^{D0}) - \alpha^2} = \sqrt{Q_1 Q_2 Q_3 Q_6 - \alpha^2} \,.
\end{eqnarray}
%

\subsubsection{Comparison with known solutions}

To the best of our knowledge, the {\it rotating} non-BPS  extremal black string solution of the $STU$ model derived above represents an original result. However, if one sets to zero the angular momentum parameter $\alpha$, our solution reduces to a non-rotating non-BPS black string which was already known in the literature. It was first found in \cite{Gimon:2007mh} by Gimon, Larsen and Simon (GLS) in terms of a four dimensional black hole, and then by Kim, Lindman, Palmkvist and Virmani (KLPV) in the context of five dimensional minimal supergravity in \cite{Kim:2010bf}. 

The map between our solution (with $\alpha=0$) and the one of \cite{Gimon:2007mh}, section 4.2, 
is straightforward, and is given by the  following identifications
\begin{eqnarray}
 \sqrt{2}H^i_{\text \tiny (GLS)} = L_I \,, \quad H_{0 {\text \tiny (GLS)}} = V \quad \mathrm{and} \quad 
 B_{\text \tiny (GLS)}= m_0\,.  
\end{eqnarray}

The solution in \cite{Kim:2010bf}, section 6.1.3, is given in conventions different from ours, and the map with our solution is a bit more involved. First of all, the paper \cite{Kim:2010bf} works in 5D minimal supergravity: our solution is easily reduced to that frame by taking all the D4 charges to be equal: $Q_1=Q_2=Q_3=Q$ (and hence $L_1=L_2=L_3=L$). Moreover, by comparing our definition of the asymptotic charges, explained before \eqref{chgesBSTR}, with the one of  \cite{Kim:2010bf}, one sees that the two differ by $\int d(B^{(2)}\wedge C^{(3)})$ , for the D2 charge, and by $\int d(B^{(2)}\wedge C^{(5)}) + d(B^{(2)} \wedge B^{(2)} \wedge C^{(3)}) $, for the D0-charge. This difference becomes irrelevant if the B-field vanishes asymptotically: this is already the case for the KLPV solution, but not for ours. We thus need to cancel the asymptotic value of our B-field by the large gauge transformation
\begin{eqnarray}
 B_I \to B_I - m_0 \,, \quad I=1,2,3 \,.
\end{eqnarray}
Notice that, as only $dB^{(2)}$ appears in our definition of the RR field strengths , this transformation does affect our values of the charges, and we can continue to use the values listed in \eqref{chgesBSTR}. We can now compare the asymptotic charges of our solution with the ones of  KLPV. This leads to the following identification: 
\begin{eqnarray}
\!\!\! Q^{D4} = Q = {q_{\text \tiny (KLPV)} \over 2} \,, \, Q^{D2} = 2 m_0 Q = Q_{\text \tiny (KLPV)} \,, \,Q^{D0} = Q_6 + 3 m_0^2 = \frac{3}{2}(q-\Delta)_{\text \tiny (KLPV)},
\end{eqnarray}
which implies that $L=V_{\text \tiny (KLPV)}$, and $e^{-4U} = g_{\text\tiny (KLPV)}$. Using this map, and taking into account the shift of the B-field, one can verify that both the metric and the gauge fields of our solution and of the one in \cite{Kim:2010bf} exactly match.

Let us finish this subsection with a small remark on the charges \eqref{chgesBSTR} of the black string. Since the D2 charges are proportional to the parameter $m_0$, which gives the asymptotic value of the B-field, one might think that they are only an effect of having a non-zero Wilson line of the B-field at infinity. We have seen, however, that the D2 charges persist even after canceling this Wilson line. One should thus think of the D2 charges as real, intrinsic charges, arising from the interactions between the other charges and the fluxes of the solution.

\subsection{A new look at Israel--Wilson spaces} \label{IW}
We have until now looked at the solutions generated by 6 T-dualities on almost-BPS solutions with vanishing D4 fluxes ($K_I=0$). The next simplest case to consider is when the starting solution has one non-vanishing D4 flux ($K_3 \neq 0$, $K_1=K_2=0$). The case with generic $K_3$ will produce solutions of a kind already considered in \cite{Bena:2009fi}, characterized by having a 4-dimensional base metric given by a Israel-Wilson space. The particular case with $K_3=1$, which we discuss in the last subsection, gives a new class of non-BPS solutions: as the almost-BPS solutions, they are built on a Gibbons-Hawking space, but their magnetic fluxes, warp factors and angular momentum vector satisfy a system of differential equations which is different from the one \eqref{nonBPSeq} describing almost-BPS solutions.

\subsubsection{Israel--Wilson spaces}
Consider an almost-BPS solution with $K_3 \neq 0$, $K_1=K_2=0$. In this case, one enjoys the drastic simplification that all the warp factors $Z_I$'s are harmonic
\begin{eqnarray}
Z_I = L_I\,,
\end{eqnarray}
 and only $\mu$ cannot be written, in general, in a closed form but is given by the solution of
\begin{eqnarray}
\label{eqmuIW}
 d \star d (V \mu) = d (V L_3 ) \wedge \star d K_3 \,. 
\end{eqnarray}
Let us look at the geometry obtained by applying 6 T-dualities on this solution, referring to the results of section \ref{galalBPS}. The geometry of the 4D base is controlled by the function $F^2$ defined in \eqref{defF}. In this case it simplifies to a perfect square, 
\begin{eqnarray}
 F^2=(K_3 L_3)^2\,.
\end{eqnarray}
Hence, if we rename
\begin{eqnarray}
\widetilde{V}_+=K_3\,,\quad  \widetilde{V}_-=L_3
\end{eqnarray}
the 4D base has the form 
\begin{eqnarray}
 &&ds^2_4 = \widetilde{V}^{-1}(d\psi+\widetilde{A})^2+ \widetilde{V} ds^2_3\,, \\[2mm] \nonumber
&&\widetilde{V} = \widetilde{V}_+\widetilde{V}_- \,,   \quad \star d \widetilde{A} = \widetilde{V}_- d\widetilde{V}_+ - \widetilde{V}_+ d\widetilde{V}_- \,,
\end{eqnarray}
where the equation for the vector $\widetilde{A}=v_0$  is the simplification of \eqref{defv0} for  $K_1=K_2=0$. This is the form of a  Israel--Wilson metric. In \cite{Bena:2009fi}, it was found that one can indeed find non-BPS 11-dimensional solutions based on Israel--Wilson spaces. 
It is interesting to compare the solution produced here by T-duality with the general class of Israel-Wilson type solutions of  \cite{Bena:2009fi}. Let us then look at the remaining quantities 
describing the 11-dimensional metric and 3-form gauge field, that we derive from the results of 
section \ref{galalBPS}:
\begin{eqnarray}
\label{newmetricdatas}
 \widetilde{Z}_1 &=& \frac{L_2}{K_3} \,, \quad \widetilde{Z}_2 = \frac{L_1}{K_3} \,, \quad \widetilde{Z}_3 = \frac{L_1 L_2}{L_3 K_3} + V (K_3 -2 \frac{\mu}{L_3}) \,, \nonumber \\
 \widetilde{\mu} &=& -\frac{L^3}{(L_3 K_3)^2} + \frac{V \mu}{L_3 K_3} \,,\nonumber\\
 \widetilde{W}_1 &=& = -\frac{L_2}{K_3} \,, \quad \widetilde{W}_2 = -\frac{L_1}{K_3} \,, \quad \widetilde{W}_3  = \frac{L_1 L_2}{L_3 K_3} + V (K_3 -2 \frac{\mu}{L_3}) \,,\nonumber \\  
 \widetilde{P}_1 &=& \frac{L_1}{K_3 L_3} \,, \quad \widetilde{P}_2 = \frac{L_2}{K_3 L_3} \,, \quad \widetilde{P}_3 = -\frac{1}{K_3} \,,\quad  \widetilde{w}^I = -v_I\,.
\end{eqnarray}
We observe that, though the new warp factors $\widetilde{Z}_I$'s are not given by harmonic functions, they are still equal, up to sign, to the electric components of the gauge-fields, given by $\widetilde{W}_I$. More precisely one has $\widetilde{W}_1 = - \widetilde{Z}_1$, $\widetilde{W}_2 = - \widetilde{Z}_2$ and $\widetilde{W}_3 =  \widetilde{Z}_3$. The sign mismatch 
in the first two identities has no physical meaning, and can be remedied, for example, by  renaming the coordinates $(y_1^1, y_2^1) \to (-y_1^1, -y_2^1)$, so as to flip the sign of the gauge field components $C^{(3)}_1$ and $C^{(3)}_2$. This coordinate redefinition results in the changes
\begin{eqnarray}
 \widetilde{W}_I \to - \widetilde{W}_I   \,, \quad  \widetilde{P}_I \to -\widetilde{P}_I \quad {\rm{and}} \quad \widetilde{w}^I \to -\widetilde{w}^I \quad \mathrm{for} \quad I=1,2\,.  
\end{eqnarray}
In the new coordinates the solution is in the form of the ``floating brane'' ansatz considered in \cite{Bena:2009fi}. In Appendix C, we check that the solution that we obtain here indeed solves the linear system of equations characterizing the family of geometries found in \cite{Bena:2009fi}.

\subsubsection{New ``almost-BPS'' solutions}

The previous class of geometries assumed a generic function $K_3$. If one considers the particular sub-case with $K_3=1$, or, in other words, in one starts from an almost-BPS geometry with trivial magnetic charges but a non-zero value for the Wilson line of $B^{(3)}_3$, the family of geometries one finds after 6 T-dualities has a quite different structure than the one discussed in the previous subsection.

For $K_3=1$ the function $\widetilde{V}_+=K_3$ that describes the Israel-Wilson base space becomes a constant, and the Israel-Wilson metric reduces to a simpler Gibbons-Hawking metric, 
whose associated harmonic function is $\widetilde{V}=\widetilde{V}_- = L_3$:
\begin{eqnarray}
 \widetilde{V} = L_3 \,, \quad \star d \widetilde{A} = -d \widetilde{V} \,.
\end{eqnarray}
As indicated by the relation above, the orientation of this Gibbons-Hawking space is negative. 
Despite the fact that this solution is built on a Gibbons-Hawking space with negative orientation, it is {\it not} in the almost-BPS class of solutions. To see this we should look at the rest of the geometry, which is encoded in the functions:
\begin{eqnarray}
 \widetilde{Z}_1 &=& L_2\,, \quad \widetilde{Z}_2 = L_1 \,, \quad \widetilde{Z}_3 = V -\frac{2 M}{L_3} + \frac{L_1 L_2}{L_3} \,,  \nonumber \\
 \widetilde{P}_1 &=& -\frac{L_1}{L_3} \,, \quad \widetilde{w}^1 = v_1 \,, \quad \widetilde{P}_2 = -\frac{L_2}{L_3} \,, \quad \widetilde{w}^2 = v_2 \,, \\ \nonumber 
\widetilde{P}_3 &=& -1 \,, \quad \widetilde{w}^3 = - v_3 \,, \\[2mm] \nonumber
 \widetilde{\mu} &=& \frac{M}{L_3} - \frac{L_1 L_2 }{L_3^2} \,, \quad \widetilde{\omega} = \omega \,.
\end{eqnarray}
From the form of $\widetilde{P}_I$ and $\widetilde{w}^I$, one sees that the field strengths of the dipole gauge fields $\tilde a_I = \tilde P_I (d\psi+\tilde A) + \tilde w^I$ do not have definite self-duality with respect to the Gibbons-Hawking base metric. This is unlike the almost-BPS solutions, which are characterized by having self-dual dipole field strengths. Despite this fact these new type of non-BPS solutions still menage to solve the full equations of motion.
 
We have thus seen that the action of 6 T-dualities on almost-BPS solutions with trivial magnetic charges but non-trivial values for the B-field Wilson lines produces a {\it new} class of non-BPS solutions based on Gibbons-Hawking base spaces.

\section{Generalisation to the full U-duality group}

As we mentioned in the introduction, our configurations can also be seen as solutions to a specific N=2 4-dimensional supergravity theory, with 3 vector multiplets parameterizing the $STU$ model.
We can therefore use the 4-dimensional U-duality group to generate new solutions starting from the ones following from our original ansatz, satisfying (\ref{BPSeq}) in the BPS case and (\ref{nonBPSeq}) in the non-BPS case.
U-duality is a generalization of the electric--magnetic duality symmetry that maps Bianchi identities and equations of motions of the vector fields among them, leaving the other equations of motion untouched.
This means that it maps solutions of the equations of motion to new solutions of the same system of equations.

The generic action of the duality group is given by a symplectic rotation matrix acting on the gauge field strengths $F^\Lambda$ and their duals $G_{\Lambda}$, defined as
\begin{equation} 
	F^\Lambda = d A_4^\Lambda, \quad G_\Lambda = {\cal R}_{\Lambda \Sigma} F^\Sigma-{\cal I}_{\Lambda \Sigma} *_4 F^\Sigma  = d A_{4\, \Lambda},
\end{equation}
while at the same time the scalar fields are redefined by an action of the group of isometries of the scalar manifold:
\begin{equation}
	\left(\begin{array}{c}
	F^\Lambda \\ G_{\Lambda}
	\end{array}\right) \to  S \left(\begin{array}{c}
	F^\Lambda \\ G_{\Lambda}
	\end{array}\right), \qquad \delta_S z^I = \xi^I(S, z).
\end{equation}
We point out that the definition of the dual curvatures $G_{\Lambda}$ is consistent with the fact that the complex symplectic vector constructed with the gauge field-strengths transforms exactly like the symplectic sections, i.e.~
\begin{equation}
	\left(\begin{array}{c}
	F^{-\Lambda} \\ G^-_{\Lambda}
	\end{array}\right) = 
	\left(\begin{array}{c}
	F^{-\Lambda} \\ {\cal N}_{\Lambda \Sigma}F^{-\Sigma}
	\end{array}\right),
\end{equation}
with $F^{\pm} \equiv \frac12 (F \mp i *_4 F)$.
For the $STU$ model this implies that $S \in {\rm SU}(1,1)^3 \subset {\rm Sp}(8,{\mathbb R})$, which can be represented by a 9-parameter matrix \cite{Behrndt:1996hu,Toldo}
\begin{equation}
	S = {\cal STU},
\end{equation}
where
\begin{equation}
	{\cal S} = \left(\begin{array}{cccccccc}
		d_1 & c_1 &&&&&& \\
		b_1 & a_1 &&&&&& \\
		&&d_1 &&&&&c_1 \\
		&&& d_1 &&& c_1 & \\
		&&&& a_1 & -b_1 && \\
		&&&& -c_1 & d_1 && \\
		&&& b_1 &&& a_1 & \\
		&& b_1 &&&&& a_1
	\end{array}\right),
\end{equation}
\begin{equation}
	{\cal T} = \left(\begin{array}{cccccccc}
		d_2 &&c_2&&&&& \\
		& d_2 &&&&&& c_2\\
		b_2 &&a_2 &&&&& \\
		&&& d_2 &&c_2&& \\
		&&&& a_2 &&-b_2& \\
		&&&b_2&& a_2 && \\
		&&&&-c_2&&d_2 & \\
		&b_2 &&&&&& a_2
	\end{array}\right),
\end{equation}
\begin{equation}
	{\cal U} = \left(\begin{array}{cccccccc}
		d_3 &&&c_3&&&& \\
		& d_3 &&&&&c_3& \\
		&&d_3 &&&c_3&& \\
		b_3&&& a_3 &&&& \\
		&&&& a_3 & && -b_3\\
		&&b_3&&& a_3 && \\
		&b_3&&&&& a_3 & \\
		&&&&-c_3&&& d_3
	\end{array}\right),
\end{equation}
and the parameters $a_I$, $b_I$, $c_I$ and $d_I$ satisfy $ a_I d_I - b_I c_I = 1$ for each $I$ and act as an SU(1,1) action on the scalars
\begin{equation}
	z^I \to \frac{a_I z^I + b_I}{c_I z^I+d_I} \qquad \hbox{(no sum).}
\end{equation}
We can use a more compact notation by introducing three matrices $M_I$, where
\begin{equation}
	M_I = \left(\begin{array}{cc}
	a_I & b_I \\
	c_I & d_I
	\end{array}\right).
\end{equation}
The combined action of the $STU$ duality group on a generic symplectic vector $(V^\Lambda,V_{\Lambda})$ can be obtained by introducing the transformation
\begin{equation}
	\label{duala}
	a_{abc}' = (M_1)_a{}^d (M_2)_b{}^e (M_3)_c{}^f\, a_{def},
\end{equation}
where 
\begin{equation}
	\begin{array}{llll}
	a_{222} = V^0,	 & a_{211} = V_1, &  a_{121} = V_2,  &  a_{112} = V_3,  \\[2mm]
	a_{111} = -V_0,	 & a_{122} = V^1, &  a_{212} = V^2,  &  a_{221} = V^3.  
	\end{array}
\end{equation}
It interesting to notice that the quantity $I_4(H^\Lambda,H_{\Lambda})$ defined in (\ref{I4def}) is invariant under the  application of a generic duality transformation.
The outcome of this discussion is that if we apply a generic duality transformation discussed above to a solution of the 4-dimensional system of equations of motion we generate a new solution.

Given the structure of the gauge vector potentials and their duals in the ansatz we employed in this paper (\ref{ALup})--(\ref{ALdown}), we can see that the action of the U-duality group can be realized as a matrix rotation by $S = {\cal STU}$ on the symplectic vector of potentials $(\chi^\Lambda, \psi_{\Lambda})$ and of 1-forms $(w^\Lambda, v_{\Lambda})$.

\subsection{The BPS case} 

Using now the simple procedure described previously, we can construct the most general BPS U-duality invariant solution by applying the duality transformation (\ref{duala}) to the scalar fields and gauge potentials defined in (\ref{chi0})--(\ref{stardv}).
Once more, the outcome can be entirely expressed in terms of harmonic functions
\begin{eqnarray}
	\star d w^{0} &=& \sqrt2 \left[ \frac{C_{IJK}}{2} \left(c_I c_J d_K dH_K +d_I d_J c_K dH^K\right)-c_1 c_2 c_3 dH_0  +d_1 d_2 d_3 dH^0 \right], \\[2mm]
	\star d w^I &=& \sqrt2 \left[ \frac{C_{IJK}}{2} (-a_I  c_J c_K dH_0+b_I  d_J d_K dH^0) + C_{IJK} d_J c_K(a_I dH_J +b_I dH^K)  \right.  \nonumber \\[3mm]
	&& \left.+ \frac{C_{IJK}}{2}(a_I d_J d_K dH^I + b_I c_J c_K dH_I)\right],
\end{eqnarray}
where the last one is summed over $J$ and $K$, but not over $I$, and, similarly,
\begin{eqnarray}
	\widetilde  \chi^{0} &=& \frac{C_{IJK}}{2} \left(c_I c_J d_K \psi_K +d_I d_J c_K \chi^K\right) - c_1 c_2 c_3 \psi_0+d_1 d_2 d_3 \chi^0 , \\[3mm]
	\widetilde  \chi^I &=& \frac{C_{IJK}}{2} (-a_I  c_J c_K \psi_0+b_I  d_J d_K \chi^0) + {C_{IJK}} d_J c_K(a_I \psi_J + b_I \chi^K)  \nonumber \\[3mm]
	&& + \frac{C_{IJK}}{2}(a_I d_J d_K \chi^I + b_I c_J c_K \psi_I),
\end{eqnarray}
where the vector $(\chi^\Lambda,\psi_{\Lambda})$ has beed constructed in (\ref{chi0})--(\ref{psiI}).

The special case of 6 T-dualities can be obtained by taking $a_I = d_I = 0$ and $b_I =-c_I = -1$, 
\begin{equation}
	M_1 = M_2 = M_3 = \left(\begin{array}{cc}
	0 & -1 \\
	1 & 0
	\end{array}\right),
\end{equation}
which reduces to 
\begin{equation}
	S = \left(\begin{array}{cc}
	0 & -1_4 \\
	1_4 & 0
	\end{array}\right)
\end{equation}
mapping
\begin{equation}
	\widetilde H^\Lambda =- H_\Lambda, \qquad \widetilde H_\Lambda = H^\Lambda
\end{equation}
and therefore
\begin{equation}
	z^I \to -\frac1{z^I}.
\end{equation}
In fact, the harmonic functions also define a symplectic vector $(H^\Lambda, H_{\Lambda})$, which rotates covariantly under duality transformations, as can be deduced from the differential relations between the harmonic functions and the vector fields.
This is exactly the same conclusion we reached in section \ref{sec:BPSdualities}.
From the analysis presented here, however, we also see that also the generic U-duality transformation leaves the original ansatz invariant, acting as a simple exchange of harmonic functions.
We will see that this is not the case for non-BPS solutions anymore.

\subsection{The non-BPS branch} 

Single centre non supersymmetric black holes in the $STU$ model are given in terms of 4 harmonic solutions and can be generated from the seed solution we reviewed in section \ref{sub:almostBPS}.
Following the procedure outlined in the previous section and already mentioned in \cite{Gimon:2007mh}, we can obtain this general single centre solution starting from a simple seed solution with only 4 charges and a non-trivial expectation value for the axion fields at infinity and acting on it with the U-duality group.
In fact, in the $STU$ model, the most general single centre configuration is specified by 5 parameters out of the 8 charges and 6 real asymptotic values of the scalar fields.
The remaining 9 parameters can be generated by an arbitrary SU(1,1)$^3$ duality transformation.

More generally, whenever the scalar manifold of the 4-dimensional supergravity theory is given by a coset space G/H, the U-duality group is identified with the isometry group G. 
Since the scalar fields parameterize the coset, we can always fix their asymptotic value at infinity to an arbitrary number by means of a non-compact transformation in $G/H$.
At this point, by using the remaining compact generators we leave the scalars invariant, but we can still transform the charges.
Hence the number of necessary parameters are the number of the charges minus the number of generators of the compact subgroup that have a non-trivial action on the same charges.
In the simple case at hand, this tells us that we need 5 parameters in total.

The result is the configuration obtained by acting with the most general duality matrices ${\cal S}$, ${\cal T}$ and ${\cal U}$ generating ($S = {\cal STU}$)
\begin{equation}
	\star dw^\Lambda =  S^\Lambda{}_0  \,dV+ S^{\Lambda I}\,  dZ_I
\end{equation}
and 
\begin{equation}
	\widetilde \chi^\Lambda =S^\Lambda{}_\Sigma \, \chi^\Sigma+S^{\Lambda \Sigma}\, \psi_\Sigma,
\end{equation} 
where $\chi^\Lambda$ and $\psi_{\Lambda}$ were given in (\ref{chisingle}) and (\ref{psisingle}).
The outcome gives the solution already presented in \cite{Bellucci:2008sv}.

An analogous operation is possible in the case of multi-center solutions, where we also associate  $P_I$ to harmonic functions ($K_I$ in \cite{Bena:2009ev}).
The outcome is that the most general solution is given by the scalar fields
\begin{equation}
	z^I  = \frac{b_I\, VZ_I + a_I (-V \mu + V K_I Z_I - i \, e^{-2U})}{d_I\, V Z_I + c_I (-V \mu + V K_I Z_I - i \, e^{-2U})}
\end{equation}
and by the vector potentials following from
\begin{eqnarray}
	\chi^0 &=& e^{4U}\left[\mu V^2  (d_1 + c_1 K_1)(d_2 + c_2 K_2)(d_3 + c_3  K_3)-c_1 c_2 c_3 Z_1 Z_2 Z_3 \right. \\[3mm]
	&&\left.+ \mu V \,\frac{C_{IJK}}{2} c_I c_J Z_K(d_K + c_K K_K)  -V \frac{C_{IJK}}{2} c_I Z_JZ_K (d_J + c_J K_J)(d_K + c_K K_K)\right], \nonumber\\[3mm]
	\chi^I &=& e^{4U}\sum_{J,K}\frac{C_{IJK}}{2} \left[\mu V^2 (b_I+a_I K_I) (d_J + c_J K_J)(d_K + c_K K_K)\right.\nonumber\\[3mm]
	&&\left.+\mu V  c_J c_K (b_I + a_I  K_I) Z_I +2 \mu V a_I c_J (d_K + c_K K_K) Z_K\right.\\[3mm]
	&& \left. - 2 V (b_I+ a_I K_I)(d_J + c_J K_J) c_K  Z_I Z_J \right. \nonumber \\[2mm]
	&&\left. -V a_I Z_J Z_K (d_J + c_J J_J)(d_K + c_K K_K)-a_I c_J c_K Z_I Z_J Z_K  \right],\nonumber	
\end{eqnarray}
and
\begin{eqnarray}
  \star dw^0 &=&  
  (d_1 + c_1 K_1)(d_2 + c_2 K_2)(d_3 + c_3  K_3) dV\nonumber \\[2mm]
  && +\frac{C_{IJK}}{2} (d_I + c_I K_I) dZ_I  c_J c_K -\frac{C_{IJK}}{2} c_I dK_I V (d_J+ c_J K_J) (d_K+ c_K K_K)\nonumber \\[2mm]
  &&-\frac{C_{IJK}}{2} c_I dK_I Z_I c_J c_K , \\[3mm]
  \star dw^I &=& \frac{C_{IJK}}{2}\left[ dZ_I (b_I + a_I K_I)c_J c_K +2 a_I d Z_J (d_J + c_J K_J) c_K \right.\nonumber\\[3mm]
  &&\left.+ (b_I +a_I K_I) (d_J+ c_J K_J) (d_K+ c_K K_K) \,dV+\right. \nonumber \\[3mm]
  &&\left. - V a_I dK_I \,(d_J+ c_J K_J) (d_K+ c_K K_K) - a_I Z_I dK_I c_J c_K \right.\nonumber \\[3mm]
  &&\left. - 2 V (b_I + a_I K_I) (d_J+ c_J K_J) \,c_K dK_K -2 a_I c_J  Z_J dK_J c_K\right].
\end{eqnarray}

Since the result reported in this section gives the most general non-BPS configuration that can be obtained by acting with the full U-duality group on the original solution obtained by employing the ansatz of section \ref{sec:thesetup}, the solution obtained by a combination of 6 T-dualities with the spectral flow presented in (\ref{4Dnewmet}) should be a special subcase of the one presented here.
This is indeed the case, provided we identify the duality matrices with
\begin{equation}
	M_I = \left(\begin{array}{cc}
	1 & 0 \\
	\gamma_I & 1
	\end{array}\right) = \left(\begin{array}{cc}
	0 & 1 \\
	-1 & 0
	\end{array}\right)\left(\begin{array}{cc}
	1 & -\gamma_I \\
	0 & 1
	\end{array}\right)\left(\begin{array}{cc}
	0 & -1 \\
	1 & 0
	\end{array}\right),
\end{equation}
which is the combination of 6 T-dualities, an axion shift ($B \to B - \gamma_I dT_I$) and again 6 T-dualities back.

\section{Conclusion}

In this paper we have studied a class of extremal non-supersymmetric solutions of the $STU$ model, or equivalently of eleven-dimensional supergravity compactified on $S^1\times (T^6/Z_2 \times Z_2)$. We have shown how, starting from a solution inside the ``floating brane" Ansatz, one can use the U-duality group to generate a general extremal non-supersymmetric solution\footnote{We recall that we only studied here extremal Reisner-Nordstr\"om-like solutions, i.e. solutions with a flat three-dimensional base. We leave the study of extremal Kerr-like solutions for future investigations.}. 

This has been done in two ways: from the point of view of the ten-dimensional T-dualities, and from the one of the U-duality group of the $STU$ model in four dimensions. This not only allows us  to build an explicit map between the two equivalent formalisms, but also gives us complementary physical insights. Indeed, while the four-dimensional approach is computationally more efficient, the ten-dimensional one allows us to keep track of the D-brane interpretation of the solutions. 

Analyzing the action of T-duality in ten dimensions also led us understand the status of spectral flow transformations within the U-duality group. While it was clear that spectral flow transformations were in the group, how they related to the other ones was not evident. 
Spectral flow is originally defined by conjugating a rotation in $S^1\times (T^6/Z_2 \times Z_2)$ by the U-duality transformation relating the M2-M2-M2 to the D1-D5-P frame \cite{Bena:2008wt}. 
In these terms, spectral flow appears as a complicated, and computationally impractical, operation. We have shown here that spectral flow has a simpler realization as the conjugation of a large gauge transformation shifting the axions by the action of 6 T-dualities on $T^6$.
 In the light of how spectral flows have already been useful in may different contexts \cite{Bena:2009fi, Balasubramanian:2000rt,Maldacena:2000dr, Lunin:2004uu, Giusto:2004id, Ford:2006yb, AlAlawi:2009qe},  
 we think that this result might be fruitful for the construction of new classes of solutions.

One should emphasize that the role dualities play in exploring the space of non-supersymmetric solutions is far more relevant than for BPS solutions. In the BPS case, indeed, one knows the general class of solutions, which is defined by the system \eqref{BPSeq}, and U-duality transformations transform solutions within this class. On the other hand the system  \eqref{nonBPSeq} only defines a subset of the full family of extremal non-BPS solutions, which is not closed under the U-duality group. The main motivation of the present paper is that acting with U-duality on the solutions in \eqref{nonBPSeq} allows us to construct many new solutions, 
such as black strings, the underrotating Rasheed--Larsen black hole \cite{Rasheed:1995zv, Larsen:1999pu} or solutions based on an Israel-Wilson space. 
It would obviously be very interesting to have independent,  first-principle, ways to construct general extremal non-BPS solutions: this would require to understand how to implement the extremality condition, thus generalizing the ``floating brane'' Ansatz, and how to reduce the general Einstein's equations to a first order system of equations describing generic extremal solutions.

\bigskip

\bigskip
\section*{Acknowledgments}

\noindent We would like to thank Iosif Bena for collaboration at the early stages of this project and for the numerous helpful discussions and suggestions. We would also like to thank Guillaume Bossard, Maria Rodriguez, Amitabh Virmani and especially Chiara Toldo for many interesting discussions. C.R.~would like to thank the IPhT, CEA Saclay and the university of Genova for kind hospitality while this work was completed.
The work of G.D.~is supported in part by the ERC Advanced Grant no. 226455, \textit{``Supersymmetry, Quantum Gravity and Gauge Fields''} (\textit{SUPERFIELDS}), by the Fondazione Cariparo Excellence Grant {\em String-derived supergravities with branes and fluxes and their phenomenological implications} and by the European Programme UNILHC (contract PITN-GA-2009-237920). The work of C.R.~is supported in part by the ANR grant 08-JCJC-0001-0 and by the ERC Starting Independent Researcher Grant 240210 - String-QCD-BH.

\smallskip
%

\section*{Appendix A. The reduction to 4 dimensions}
\appendix
\renewcommand{\theequation}{A.\arabic{equation}}
\setcounter{equation}{0} \addcontentsline{toc}{section}{Appendix A. The reduction to 4 dimensions}
\label{reductionto4d}

We report here some details on the identification of the fields of 11-dimensional supergravity with the ones of 4-dimensional supergravity for the ansatz used in this paper.
Some additional details of these identifications can be found in \cite{Toldo}.

\subsection*{A.1 The reduction from 11d to 5d}

The reduction from 11 to 5 dimensions follows from the reduction ansatz of the metric
\begin{equation}
ds_{11}^2 = ds_{5d}^2 +  X^1 (dy_{1,1}^2+dy_{1,2}^2)+ X^2 (dy_{2,1}^2+dy_{2,2}^2)+X^3 (dy_{3,1}^2+dy_{3,2}^2)	
\end{equation} 
and of the 3-form
\begin{equation}
	A^{(3)} = \sum_{I=1}^3 A_5^I \wedge dT_I,
\end{equation}
inserted in the Lagrangian
\begin{equation}
	2 \kappa_{11}^2 \, S = \int \sqrt{-g_{11}} \left(R_{11} - \frac{1}{48}F_{\underline{MNPQ}}F^{\underline{MNPQ}}\right) - \frac16 \int F^{(4)} \wedge F^{(4)} \wedge A^{(3)}.
\end{equation}
The ansatz we took for the metric is such that in the reduction to 5 dimensions we keep only the vector multiplet vectors and scalars, while we set to zero all the hypermultiplet (we froze the overall volume and all the complex structure moduli of the tori).
For an internal manifold of fixed volume ($g_6 = 1$) we therefore get the reduced Ricci Einstein--Hilbert action 
\begin{equation}
	2 \kappa_{11}^2 \, S = \int_{M_5 \times Y_6} \sqrt{-g_{11}} R_{11} = \int_{M_5}  \sqrt{-g_{5}} \left(R_{5} - \frac14 g^{li} g^{jk}\partial_M g_{ij}\partial^M g_{kl}\right),
\end{equation}
which in our case becomes ($\kappa_{5}^2 = \kappa_{11}^2/{\rm Vol}(Y_6)$)
\begin{equation}
	2 \kappa_{5}^2 \, S_5 \!= \!\! \int_{M_5} \!\!\!\! \sqrt{-g_{5}} \left( \! R_{5} \! - \! \frac12 \left[\frac{1}{(X_1)^2}\partial_M X^1\partial^M X^1 \! + \frac{1}{(X_2)^2}\partial_M X^2\partial^M X^2 \! + \frac{1}{(X_3)^2}\partial_M X^3\partial^M X^3\right] \! \right).
\end{equation}

The full reduced action is 
\begin{equation}
	2 \kappa_{5}^2 \, S_5 \! = \! \! \int_{M_5} \!\!\!\! \sqrt{-g_{5}} \left( \! R_{5} \! - \! G_{AB} \partial_M X^A \partial^M X^B \! - \frac12 G_{AB} F^A_{MN} F^{B\,MN} \right) - \frac13 \int C_{ABC} F^A \wedge F^B \wedge A_5^C,
\end{equation}
where 
\begin{equation}
	C_{123} = \frac12, \quad {\cal K} = \frac16 C_{ABC} X^A X^B X^C = \frac12,
\end{equation}
and
\begin{equation}
	G_{AB} = - \frac12 (\partial_A \partial_B \log {\cal K})|_{{\cal K} = 1/2}.
\end{equation}

\subsection*{A.2 The reduction from 5d to 4d} 

The reduction ansatz from 5d to 4d follows from 
\begin{equation}
	ds^2_{5d}= \Delta^2 ds_{4d}^2 + \frac{1}{\Delta^4} \left(d \psi - A^0\right)^2,
\end{equation}
for the metric, 
\begin{equation}
	A_5^I = A_4^I + C^I (d \psi - A^0)
\end{equation}
for the vectors and 
\begin{equation}
	z^I = C^I - i\, \frac{X^I}{\Delta^2}
\end{equation} 
for the scalars.
In this way we reduce to the 4d $STU$ model, for a lagrangian of the form
\begin{equation}
	{\cal L}_4 = \frac12\, R - g_{i \bar \jmath}\partial_\mu z^i \partial^\mu \bar z^{\bar \jmath} + \frac18\, {\cal I}_{\Lambda \Sigma} F_{\mu\nu}^\Lambda F^{\Sigma\,\mu\nu} + \frac18\, {\cal R}_{\Lambda \Sigma} F_{\mu\nu}^\Lambda (*_4 F)^{\Sigma\,\mu\nu},
\end{equation}
with $(*_4 F)_{\mu\nu} = \frac{1}{2} \sqrt{-g}\, \epsilon_{\mu\nu\rho\sigma} F^{\rho \sigma}$.
Relabelling the scalar fields as $z^I = \{S = \sigma - i\, s, T = \tau - i\, t, U = \upsilon - i\, u\}$, the metric of the scalar $\sigma$-model $g_{I\bar J}$ follows from the K\"ahler potential
\begin{equation}
  K = - \log(8 \, stu),
\end{equation}
the gauge kinetic couplings are
\begin{equation}
	{\cal I} = - stu\, \left(\begin{array}{cccc}
	1+\frac{\sigma^2}{s^2}+\frac{\tau^2}{t^2}+\frac{\upsilon^2}{u^2} & -  \frac{\sigma}{s^2} & -  \frac{\tau}{t^2} & - \frac{ \upsilon}{u^2} \\[2mm]
 	-  \frac{\sigma}{s^2}& \frac{1}{s^2} & 0 & 0 \\[2mm]
	-  \frac{\tau}{t^2} & 0 & \frac{1}{t^2} & 0 \\[2mm]
	- \frac{ \upsilon}{u^2} & 0 & 0 & \frac{1}{u^2} 
	\end{array}\right).
\end{equation}
and the axionic couplings are
\begin{equation}
	{\cal R} = \left(\begin{array}{cccc}
	2 \sigma \tau \upsilon & - \tau \upsilon & - \sigma \upsilon & - \sigma \tau \\
	-\tau \upsilon & 0 & \upsilon & \tau \\
	- \sigma \upsilon & \upsilon & 0 & \sigma \\
	- \sigma \tau & \tau & \sigma & 0 
	\end{array}\right),
\end{equation}

Moreover, we are looking for a 4d metric that describes multi-center black hole configurations, hence
\begin{equation}
	ds_{4d}^2 = - {\rm e}^{2U} (dt +  \omega)^2 + {\rm e}^{-2 U} ds_3(\vec{x})^2 \,,
\end{equation}
where $ds_3(\vec{x})^2$ is the flat three-dimensional metric. By comparison of the two metrics 
\begin{equation}
	\begin{array}{rcl}
	ds_{5d}^2 &=& \displaystyle - Z^{-2} (dt +  \omega + \mu (d \psi +  A))^2 + V Zd \vec{x}^2 + \frac{Z}{V} ( d \psi +  A)^2  \\[2mm]
	&=& \displaystyle - \Delta^2 {\rm e}^{2U} (dt + \omega)^2 + \Delta^2 {\rm e}^{-2U }d\vec{x}^2 + \frac{1}{\Delta^4}(d \psi +  A + \alpha (dt + \omega))^2
	\end{array}
\end{equation}
we get that
\begin{equation}
	\alpha = - \frac{V \mu}{Z^3- V \mu^2},
\end{equation}
\begin{equation}
	\Delta^4 = \frac{Z^{2/3}V}{Z^3- V \mu^2}
\end{equation}
and
\begin{equation}
	{\rm e}^{-2U} = \sqrt{V Z^3 - V^2 \mu^2} = \sqrt{V Z_1 Z_2 Z_3 - \mu^2 V^2}.
\end{equation}
Comparing the vectors and scalars we also get the expression for the 4-dimensional vector fields:
\begin{equation}
	A_4^0 = -A + {\rm e}^{4U} \,\mu\, V^2 (dt + \vec \omega),
\end{equation}
\begin{equation}
	A_4^I = w^I - \frac{e^{4U}V}{Z_I} \left(Z_1 Z_2 Z_3-\mu V P_I Z_I\right)(dt + \vec\omega),
\end{equation}
and finally
\begin{equation}
	z^I = \frac{\left(V Z_I P_I -  V \mu\right) - i\, {\rm e^{-2 U}}}{V Z_I},
\end{equation}
where $Z_I P_I$ is not summed over $I$.

\section*{Appendix B. The computation of the dual gauge fields}
\appendix
\renewcommand{\theequation}{B.\arabic{equation}}
\setcounter{equation}{0} \addcontentsline{toc}{section}{Appendix B. The computation of the dual gauge fields}

In this Appendix, we compute the gauge potentials $C^{(5)}$ and $C^{(7)}$ dual to $C^{(1)}$ and $C^{(3)}$ given in \eqref{BPS10D} for supersymmetric solutions. We recall that the field strength are given by 
\begin{eqnarray}
H^{(3)} &=& \sum_{I=1}^3 H^{(3)}_I \wedge dT_I = \sum_{I=1}^3 dB^{(2)}_I\wedge dT_I \,, \nonumber \\
F^{(2)} &=& d C^{(1)} \,, \\ \nonumber
F^{(4)} &=& \sum_{I=1}^3 F^{(4)}_I \wedge dT_I = \sum_{I=1}^3 (dC_I^{(3)} - H^{(3)}_I \wedge C^{(1)})\wedge dT_I \,.
\end{eqnarray}
The dual fields strengths are
\begin{equation}
\label{eqF6}
F^{(6)}=-*_{10} F^{(4)} = d C^{(5)}- H^{(3)}\wedge C^{(3)}=\sum_{J,K} \Bigl({1\over 2}dC^{(5)}_{JK}-H^{(3)}_J \wedge C^{(3)}_K \Bigr)\wedge dT_J\wedge dT_K\,,
\end{equation}
\begin{equation}
\label{eqF8}
F^{(8)}=*_{10} F^{(2)}= dC^{(7)}-H^{(3)}\wedge C^{(5)} = dC^{(7)}- \sum_{I,J,K} {1\over 2} H^{(3)}_I \wedge C^{(5)}_{JK}\wedge dT_I\wedge dT_J\wedge dT_K\,.
\end{equation}
$*_{10}$ denotes the Hodge dual with respect to the string metric $ds^2_{10}$.

In order to have an explicit expression for the potentials, one now need to specify the to the particular solutions \eqref{11Dfields}-\eqref{BPSeq}:
\begin{eqnarray}
H^{(3)}_I &=& d \Bigl( {K_I\over V} - {\mu\over Z_I}\Bigr) \,, \\
F^{(2)} &=& (dt+\omega) \wedge d \Bigl( e^{4U}\,V^2\mu \Bigr) + e^{4U} \star \Bigl[ V^3 \mu Z_I d\Bigl( {K_I\over V} \Bigr) - V^3 \mu d\mu + Z^3 V dV \Bigr] \,, \\
F^{(4)}_I &=& e^{4U} (dt+\omega) \wedge \Bigl[ V Z^3 d Z_I^{-1} + {V^2 \mu \over Z_I} d\mu - V^2 \mu \, d\Bigl( {K_I\over V} \Bigr) \Bigr] \nonumber \\
&& \quad + {V\over Z_I} \star \Bigl[ Z_A d \Bigl( {K_A\over V}\ \Bigr) - d\mu \Bigr] \,, \\
*_{10} F^{(4)} &=& \Bigl \{ {Z_I\over Z^3} (dt+\omega) \Bigl[ Z_A d \Bigl( {K_A\over V} \Bigr) - d\mu \Bigr] \nonumber \\
&& \quad + \star \Bigl[ d Z_I - {V \mu Z_I\over Z^3} d\mu + {V \mu Z_I^2 \over Z^3} d\Bigl( {K_I\over V} \Bigr) \Bigr] \Bigr\} \wedge {C_{IJK} \over 2} dT_J \wedge dT_K \,, \\
 *_{10} F^{(2)} &=& \Bigl\{ (dt+\omega) \wedge \Bigl[ {\mu \, Z_I\over Z^3} d\Bigl( {K_I\over V} \Bigr)- {\mu \over Z^3} d\mu - dV^{-1}\Bigr] \nonumber \\
 && \quad - {e^{-8U} \over V^3 Z^3} \star d\Bigl( e^{4U}\,V^2\mu \Bigr) \Bigr\} \wedge dT_1\wedge dT_2\wedge dT_3\,.
\end{eqnarray}
In the non-BPS case, the computation is similar. 
One can now integrate to obtain $C^{(5)}$ and $C^{(7)}$: 
\begin{eqnarray}
C^{(5)}_{JK} &=& {\mu\over Z_J Z_K} (dt+\omega) - C_{IJK} v_I + \Bigl( {K_J\over V} - {\mu\over Z_J} \Bigr) \Bigl( {K_K\over V} - {\mu\over Z_K} \Bigr) A \nonumber \\[2mm]
&& \quad + \Bigl( {K_J\over V} - {\mu\over Z_J} \Bigr) w^K + \Bigl( {K_K\over V} - {\mu\over Z_K} \Bigr) w^J \,,
\end{eqnarray}
where
\begin{equation}
\star dv_I = d L_I \,, 
\end{equation}
and
\begin{eqnarray}
C^{(7)} \!\!&=&\!\! \Bigl\{ {e^{-4U}\over V^2 Z^3} (dt+\omega) -v_0 - \Bigl( {K_I\over V} - {\mu\over Z_I} \Bigr) v_I + \Bigl( {K_1\over V} - {\mu\over Z_1} \Bigr) \! \Bigl( {K_2\over V} - {\mu\over Z_2} \Bigr) \! \Bigl( {K_3\over V} - {\mu\over Z_3} \Bigr) A \nonumber \\[2mm]
 && \quad + {C_{IJK}\over 2} \Bigl( {K_I\over V} - {\mu\over Z_I} \Bigr) \Bigl( {K_J\over V} - {\mu\over Z_J} \Bigr) w^K  \Bigr\} \wedge dT_1\wedge dT_2\wedge dT_3 \,,
\end{eqnarray}
where
\begin{equation}
\star dv_0 = 2d M \,.
\end{equation}
Note that we have used the explicit form of the BPS solution
\begin{equation}
Z_I = L_I +  {C_{IJK}\over 2} {K_J K_K\over V}\,,\quad \mu = M + {L_I K_I\over 2 V}+ {K_1 K_2 K_3\over V^2}\,.
\end{equation}

\section*{Appendix C. Equations of motions for solutions on an Israel--Wilson base}
\appendix
\renewcommand{\theequation}{C.\arabic{equation}}
\setcounter{equation}{0} \addcontentsline{toc}{section}{Appendix
C. Equations of motions for solutions on an Israel--Wilson base}

In this Appendix, we verify that the solution presented in section \ref{IW} falls in the class of solutions found in \cite{Bena:2009fi}, based on an Israel-Wilson space. In order to do that, one has to proceed step by step. The first equation to check is the equation relating the base space with the Maxwell fields 
\begin{equation} \label{EMbaseq}
 \widehat R_{ab}  ~=~  - {\cal T}_{ab} \big( \Theta^{(3)}   \,,  \omega^{(3)}_-\big) \,.
\end{equation}
The tensor  ${\cal T}_{ab}$ is defined for any pair of 2-forms $X,Y$ by
\begin{equation}
{\cal T}_{ab} (X,Y) ~\equiv~ \frac{1}{2} \, \big( X_{ac} \, Y_{bc} ~+~ X_{bc} \, Y_{ac} \big) ~-~ \frac{1}{4} \, \delta_{ab} \, X_{cd} \, Y_{cd} \,.
\label{Tdefn}
\end{equation}
We recall that $\Theta^{(I)}= d a_I$, that $\Theta^{(3)}$ is self-dual, and that  we defined $\omega^{(3)}_-$ by\footnote{See \cite{Bena:2009fi} for details.}
\begin{eqnarray} \label{defomega3}
\big(\Theta^{(1)} - * \Theta^{(1)} \big)  &=&  2 Z_2 \, \omega^{(3)}_- \,, \qquad
\big(\Theta^{(2)} - * \Theta^{(2)} \big)  ~=~  2 Z_1 \, \omega^{(3)}_- \,.
\end{eqnarray}
Using \eqref{newmetricdatas} and \eqref{defomega3}, one has
\begin{eqnarray}
\tilde{a}_3 &=& \frac{K_+}{\widetilde{V}_+}(d\psi + \widetilde{A}) + \widetilde{w}^3 \,, \quad \star d\widetilde{w}^3 = -V_- d K_+ + K_+ d V_- \,, \\ \nonumber
\omega^{(3)}_- &=& d\left(\frac{K_-}{\widetilde{V}_-}(d\psi + \widetilde{A}) + \widetilde{w}_- \right) \,, \quad \star d\widetilde{w}_- = V_+ d K_- - K_- d V_+ \,,
\end{eqnarray}
with\footnote{We keep the notation with generic $K_+$ and $K_-$ for later conveniency.} $K_+=K_-=-1$. In order to solve \eqref{EMbaseq}, $K_+$ and $K_-$ have to be harmonic, and to verify
\begin{equation} \label{eincond}
\partial_i \Big({ K_+ \over V_+}\Big) \partial_j \Big({K_- \over V_-}\Big) =  \big(\partial_i V_+^{-1}\big) \,\big(\partial_j V_-^{-1}\big)\,.
\end{equation}
This is obviously the case here, as $K_+=K_-=-1$. The rest of the equations to be solved are 
\begin{eqnarray}
\label{systK2Z1b2}
 \star d w^2 &=& - V_- d  K_2 + K_2 d  V_-  + 2 \, V_+ V_- Z_1 d\left({K_- \over V_- } \right) \,, \nonumber \\[3mm] 
 d \star d K_2 &=& {2 \over V_-} \, d \left( V_+ V_- Z_1 \star d\left({K_- \over V_- } \right)\right) \,, \\[3mm] \nonumber
 d \star d Z_1 &=&  V_- d \star d \left({K_2 K_+ \over V_+} \right) ~-~  2 \, d \left[ Z_1 K_+  V_- \star d \left( { K_- \over V_-} \right) \right] \,,
\end{eqnarray}
plus an analogous system of equations for $w^1,K_1$ and $Z_2$, and
\begin{eqnarray}
\label{sysmuZ3omega}
  \star d \omega &=& V_+^2 d \left({V_-\over V_+}\mu \right) - V_+ V_- Z_I \, d \left( {K_I\over V_+} \right) + 2 \, Z_1 Z_2 V_+ V_- d\left( {K_- \over V_-} \right) \,, \nonumber \\[3mm]
 d \star d (V_- \mu) &=& {1 \over V_+} d \left[ V_- V_+ Z_I \star d \left({K_I \over V_+}\right) \right] - {2 \over V_+} \, d \left[ Z_1 Z_2 V_+ V_- \star d \left( { K_- \over V_-} \right) \right] \,, \\[3mm] \nonumber
d \star d Z_3 &=&  V_- d \star d  \Big({K_1 K_2 \over V_+} \Big) ~-~  2 \, d \left[ (Z_1K_1 + Z_2 K_2) V_- \star d \left( { K_- \over V_-} \right) \right] \\[3mm] \nonumber 
 && + V_+ V_- d \left( { K_- \over V_- } \right) \left[ 2 \star d \mu - Z_I \star d \left( { K_I \over V_+ } \right) + 2 \, Z_1 Z_2 \star d \left( { K_- \over V_- } \right) \right] \,.
\end{eqnarray}
Pluging in the fields given in \eqref{newmetricdatas}, and using equations \eqref{eqmuIW} and \eqref{nonBPSeq} for $\mu$ and $\omega$, it is lengthy but straightforward to check that we exactly solve these equations. In other words, we have shown that performing 6 T-dualities on a solution where only one of the magnetic charge is turned on brings us to the Israel--Wilson class of solutions discovered in \cite{Bena:2009fi}.

\section*{Appendix D. A lonely supertube}
\appendix
\renewcommand{\theequation}{D.\arabic{equation}}
\setcounter{equation}{0} \addcontentsline{toc}{section}{Appendix
D. A lonely supertube}

In the class of solution presented in section \ref{IW}, the base is an Israel--Wilson space. 
\begin{eqnarray}
 ds_4^2 &=& (V_+ V_-)^{-1} (d \psi + A)^2 + V_+ V_- ds_3^2 \,, \\ \nonumber
 \star dA &=& V_- d V_+ - V_+ d V_- \,. 
\end{eqnarray}
We know that if we assume $V_-=L_3=1$, the solutions become BPS, and should therefore be described by the BPS Ansatz given in section \ref{sec:thesetup}. We verify it here, as a consistency check. To fix the ideas, let us imagine to start from a single ``almost-BPS'' supertube in Taub-NUT space:
\begin{eqnarray}
\label{almostBPStube}
 V &=& h + \frac{Q_6}{r} \,, \quad A = -Q_6 \cos\theta d\phi \,, \nonumber \\ 
 K_1 &=& K_2 = 0 \,, \quad K_3 = k_3 + \frac{d_3}{\Sigma} \,, \quad Z_I = L_I = l_I + \frac{Q_I}{\Sigma} \,, \quad I=1,2 \\ \nonumber 
 L_3 &=& 1 \,, \quad M = m_0 + \frac{m}{\Sigma} + \frac{\widetilde{m}}{r} \,, \quad \mu = \frac{M}{V} + \frac{K_3}{2} \,.
\end{eqnarray}
In this case, the solution can be written in a closed form. As explained in \cite{Bena:2009ev}, we know that this ``almost-BPS'' supertube is secretly BPS, and therefore we expect to recover a BPS solution even after the T-dualities. We first see that for $L_3=1$, the base space is a Gibbons-Hawking space with positive orientation:
\begin{eqnarray}
 \widetilde{V} = \widetilde{V}_+ = K_3 \,, \quad \star d \widetilde{A} = d \widetilde{V} \,.
\end{eqnarray}
The remaing metric functions are
\begin{eqnarray}
 \widetilde{Z}_1 &=& \frac{L_2}{\widetilde{V}} \,, \quad \widetilde{Z}_2 = \frac{L_1}{\widetilde{V}} \,, \quad \widetilde{Z}_3 = -2 M + \frac{L_1 L_2}{\widetilde{V}} \,,  \nonumber \\
 \widetilde{P}_1 &=& -\frac{L_1}{\widetilde{V}} \,, \quad \widetilde{P}_2 = -\frac{L_2}{\widetilde{V}} \,, \quad \widetilde{P}_3 = -\frac{1}{\widetilde{V}} \,, \\ \nonumber
 \widetilde{\mu} &=& \frac{V}{2} + \frac{M}{\widetilde{V}} - \frac{L_1 L_2 }{\widetilde{V}^2}\,.
\end{eqnarray}
This is exactly the form of a BPS solution associated with the harmonic functions $\tilde K_I$, $\tilde L_I$, $\tilde M$ with
\begin{eqnarray}
\label{tubemap}
 \widetilde{K}_1 = - L_1 \,, \quad \widetilde{K}_2 = - L_2 \,, \quad \widetilde{K}_3=-1 \,, \quad \widetilde{L}_1=\widetilde{L}_2=0 \,, \quad \widetilde{L}_3=-2M \,, \quad \widetilde{M} = \frac{V}{2} \,.
\end{eqnarray}
The corresponding vector fields 
\begin{eqnarray}
 \widetilde{w}^I = v_I \,, \quad \widetilde{\omega} = \omega 
\end{eqnarray}
verify the expected relations
\begin{eqnarray}
 \star d\widetilde{w}^I = - d \widetilde{K}_I  \quad \mathrm{and} \quad \star d\omega = \widetilde{V} d \widetilde{M} - \widetilde{M} d \widetilde{V} -\frac{1}{2}( \widetilde{K}_I d \widetilde{L}_I - \widetilde{L}_I d \widetilde{K_I} ) \,.
\end{eqnarray}
This therefore checks that the solution for $L_3=1$ is supersymmetric. The analysis above is also consistent with the fact the starting  ``almost-BPS'' supertube \eqref{almostBPStube} can be recast in an explicit BPS form through the redefinitions: 
\begin{eqnarray}
 \widehat{K}_3 = 2M \,, \quad \widehat{M} = \frac{K_3}{2} 
\end{eqnarray}
These redefinitions, combined with the T-duality rules for BPS solutions  \eqref{bpst6}, exactly reproduce the relations \eqref{tubemap}.



\end{document}